\documentstyle[aps,epsfig]{revtex}

\begin{document}

\draft


\title{The $\Delta (1232)$--nucleon interaction in the $^2$H$(p,n)$ 
charge exchange reaction}

\author{C.~A. Mosbacher and F. Osterfeld}

\address{Institut f\"ur Kernphysik, Forschungszentrum J\"ulich GmbH, 
  D--52425 J\"ulich, Germany}

\date{April 11, 1997}

\maketitle


\begin{abstract}

The $^2$H$(p,n)$ charge exchange reaction at $T_p=790$ MeV is used
to study the $\Delta(1232)$--nucleon ($\Delta N$)
interaction in the $\Delta$ resonance excitation energy region.
For the $\Delta N$ potential, a meson exchange model is adopted where
$\pi$, $\rho$, $\omega$, and $\sigma$ meson exchanges are taken into
account.
The deuteron disintegration below and above pion threshold
is calculated using a coupled channel approach.
Various observables, such as the inclusive cross section, 
the quasifree $\Delta$ decay, the coherent pion production,
and the two--nucleon breakup are considered.
It is shown that these observables are influenced by the dynamical 
treatment of the $\Delta$ degrees of freedom.
Of special interest is the coherent pion decay of the $\Delta$ 
resonance which is studied by means of the exclusive reaction
$^2$H$(p,n\pi^+)^2$H.
Both the peak energy and the magnitude of the coherent
pion production cross section depend very sensitively on the strength
of the $\Delta N$ potential. The coherent pions have a peak energy
of $\omega=300$ MeV and a strongly forward peaked angular
distribution.

\end{abstract}
                                                                                 
\pacs{25.40.Kv, 25.10.+s, 14.20.Gk, 24.10.Eq}



\section{Introduction}


In recent years, inclusive and exclusive
$(p,n)$ and $( ^3He,t )$ charge exchange reactions
at intermediate energies have been proven to be excellent
probes for investigating the $\Delta$ dynamics in nuclei.
The most important observation of the inclusive charge exchange reactions
is the downward energy shift of the $\Delta$ resonance peak position
by $\approx 70$ MeV in nuclear targets (with mass number $A \geq 10$)
as compared to the proton target 
\cite{contardo86,lind87,gaarde91,gaarde96,prout96}.
Microscopic $\Delta$--hole model calculations show
that a large part of this shift is caused by the attractive
$\Delta$--nucleon ($\Delta N$) interaction in the spin--longitudinal (LO) 
channel \cite{oset82,udagawa90,delorme91,osterfeld92,udagawa94}. 
A direct signature of this interaction is provided by the measurement 
of the coherent pion decay spectrum where the pion of the $\Delta$ resonance 
decay is measured in coincidence  with the ejectile while the target nucleus 
is left in its ground state
\cite{chiba91,hennino93,oltmanns93,koerfgen94,cordoba93,cordoba95}.
These coherent pions couple strongly to the LO channel.
Due to the attractive $\Delta N$ potential in this channel,
their energy spectrum is substantially shifted downwards
relative to the $\Delta$ resonance peak position of the inclusive 
reaction.

In this paper we study the $\Delta$ excitation in the deuteron target
using the $^2$H$(p,n)$ charge exchange reaction at $T_p$=790 MeV.
The deuteron has the advantage that its wave function is well known 
and that the Fermi motion of the $\Delta N$ system can be treated
properly. Therefore the effects of the $\Delta N$ interaction
can be studied in a more direct way than in heavier nuclei.
In the past most of the studies on the deuteron have been
carried out with electromagnetic and  hadronic probes, such as
in photon--deuteron ($\gamma d$) and pion--deuteron ($\pi d$) 
scattering, leading to $NN$, $\pi d$, and $\pi NN$ final reaction 
channels. In these reactions the intermediate $\Delta N$ interaction 
plays an important role
\cite{maxwell80,leidemann87,niskanen95,wilhelm95,wilhelm96,schmidt96,kamalov97}. 
The photon excites the $\Delta$ dominantly with spin--transverse (TR) 
coupling, i. e. by the transition operator $\vec{S}\times\vec{q} \: \vec{T}$
($\vec{S}$ and $\vec{T}$ are the spin and isospin transition
operators, respectively), while the pion excites it
with spin--longitudinal ($\vec{S}\cdot\vec{q} \: \vec{T}\, )$ coupling.
Both couplings are orthogonal to each other and therefore
give different information on the $\Delta N$ interaction.
In the charge exchange reactions the target is exposed to
the virtual $\pi$ and $\rho$ meson fields produced by the $(p,n)$ 
projectile--ejectile system.
The pion--like interaction excites the LO response function
($\vec{S}\cdot\vec{q}\: \vec{T}$ coupling)
of the target while the $\rho$--meson--like interaction
excites the TR response function
($\vec{S}\times\vec{q}\: \vec{T}$ coupling).
Due to the kinematics, the virtual meson fields obey the
energy--momentum relation $\omega<q$ and thus explore the LO and TR
response functions in an $(\omega,\vec{q}\, )$ region which is
inaccessible to real pion and real photon scattering.

In the interpretation of the ($\gamma d$) and ($\pi d$) scattering
reactions, various theoretical models have been applied 
to treat the $\Delta N$ dynamics. Among these models are 
the coupled channel approach 
\cite{green,niskanen78,lee84,lee87,poepping87,pena92}
and the three--body Faddeev treatment of the $\pi NN$ system
\cite{fadeev61,blaazer95}.
In the present paper we make use of a coupled channel approach 
to describe the $\Delta N$ system in a non--relativistic framework. 
We set up a system of coupled equations for the $\Delta N$ wave function 
in configuration space. We then apply the Lanczos method 
for solving the equations.
The $\Delta N$ potential is constructed within a meson--exchange 
model. The exchanged mesons taken into account are the pion ($\pi$),  
the rho ($\rho$), the omega ($\omega$), and the sigma ($\sigma$).

The aim of the present paper is to show that the $(p,n)$ reaction 
at forward scattering angles is an interesting tool to study the
$\Delta N$ interaction in the deuteron.
The advantage of the $(p,n)$ reaction over other probes is twofold: 
First, at forward scattering angles the quasi--elastic peak cross 
section and the $\Delta$ resonance peak cross section are energetically
well separated. The $\Delta$ resonance cross section
is very large as compared to the quasi--elastic cross section
providing a well--defined $\Delta$ resonance peak in the spectrum.
Second, the $\Delta$ resonance cross section involves both a
LO and TR component (with a ratio of LO/TR = 1/2) and thus allows 
to examine the complete spin structure of the $\Delta N$ interaction.
The LO excitation is of special interest for the
coherent pion decay of the $\Delta N$ system. Due to
the LO spin structure in both the excitation and de--excitation process
the cross section for coherent pion production becomes relative large.
Because the $\Delta N$ interaction is most attractive in the 
LO channel, the coherent pion spectrum is expected to clearly exhibit 
the effects of the $\Delta N$ potential.
It turns out, indeed, that both the peak energy and the magnitude of 
the coherent pion production cross section depend quite sensitively on the
strength of this potential. Other observables, such as
the inclusive cross section, the quasifree $\Delta$ decay cross section,
and the two--nucleon breakup, are also shown to be influenced by the dynamical 
treatment of the $\Delta$ degrees of freedom.

The organization of the present paper is at follows.
In Sec.\ 2 we give a detailed account of the formulation 
and methods of calculation used in the analysis of the data.
First we present the coupled channel approach and show how to calculate
the correlated $\Delta N$ wave function in a very efficient way 
with the Lanczos method. 
Then we split the inclusive cross section into its various partial 
cross sections, as there are the contributions of coherent pion 
production, quasifree $\Delta$ decay, and two--nucleon breakup.
In Sec.\ 3 we discuss the parameters used in our model. 
We present the results of the cross section calculations and compare 
them to experimental data \cite{prout,prout94}.
Finally, in Sec.\ 4 we give a summary and conclusions.


\section{Theory}

We are interested here in the calculation of the inclusive and
exclusive cross sections for the $^2$H$(p,n)$ charge exchange reaction
in the $\Delta$ resonance energy region. Since we shall deal with
high projectile energies ($T_p = 790$ MeV), we assume that
the cross sections can be calculated within the distorted wave impulse 
approximation (DWIA). Because of the transparency of the deuteron target 
it is sufficient to calculate the distortion effects within the eikonal 
approximation.


\subsection{Inclusive cross section}

We start our formulation by writing down the formula for the inclusive 
$^2$H$(p,n)$ cross section. Using relativistic kinematics, this 
cross section is given as   
\begin{equation}
  \label{eq1}
  d\sigma = \frac{2 E_p E_d}{\sqrt{\lambda(s,M_p^2,M_d^2)}}
  \: \frac{M_p}{E_p} \frac{M_n}{E_n} \: \frac{d^3 p_n}{(2 \pi)^3 \, 2E_n} \:
  \overline{\sum} \: (2 \pi) \delta(E_p+E_d-E_n-E_f) \mid T_{fi} \mid^2 .
\end{equation}
Here the indices $p(n)$ and $d$ refer to the proton projectile 
(neutron ejectile) and the deuteron target, respectively. 
$E_f$ is the total energy of the final 
state on the target side, i.e. of all outgoing particles except the neutron 
ejectile. The c.m. momentum of the final state is fixed by momentum and 
energy conservation while an integration over the relative momenta has 
to be performed. In addition, 
an average over the initial spin orientations and a sum over the final spin 
orientations of both the projectile and the target spin states are taken.
The transition amplitude will be evaluated in the Breit frame (BF) of the 
target system, where the deuteron state is well described by the usual 
nonrelativistic wave function. 

Since we want to compare our results with experimental
data where the energy and the scattering angle of the outgoing neutron have
been measured, we have to calculate the double differential cross section
$d^2\sigma / dE_n d\Omega_n$ in the laboratory frame (lab).
From Eq.\ (\ref{eq1}) we obtain
\begin{eqnarray}
  \label{eq2}
  \left( \frac{d^2\sigma}{dE_n d\Omega_n} \right)^{\text{lab}} &=& 
  \frac{p_n^{\text{lab}}}{p_n} 
  \left( \frac{d^2\sigma}{dE_n d\Omega_n} \right)^{\text{BF}} 
  \nonumber \\ &=& 
  \frac{M_p M_n}{(2 \pi)^2} \, \frac{p_n^{\text{lab}}}{p_p^{\text{lab}}}
  \frac{E_d}{M_d} \:  
  \overline{\sum} \: \delta (E_p-E_n+E_d-E_f) \, \mid T_{fi} \mid^2 .
\end{eqnarray}
In the following, cross section results will always be given in the 
laboratory system while every other unlabeled quantity refers to the 
Breit Frame.


\subsection{The uncorrelated source function}
\label{sec:dwia}

\begin{figure}
  \caption{Impulse approximation diagrams for the $^2$H$(p,n)$ reaction.
     Here (a) shows the nucleon excitation and (b) the $\Delta$ excitation.} 
  \label{fig1}
  \begin{center}
    \epsfig{file=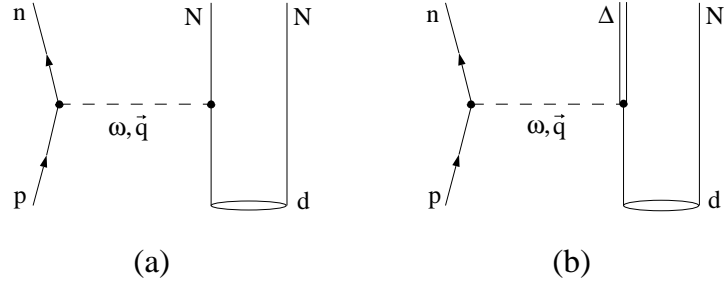, height=3.8cm}
  \end{center}
\end{figure}

In this section we discuss the excitation processes which contribute to 
$^2$H$(p,n)$ in impulse approximation. In our model, there are two 
relevant Feynman diagrams for this case which are shown in 
Fig.\ \ref{fig1}. They represent the nucleon excitation (a) and the $\Delta$ 
excitation (b) from the deuteron target.   
In first order DWIA the transition amplitude $M_{fi}$ corresponding to 
the graphs in Fig.\ \ref{fig1} is given as
\begin{equation}
  \label{eq3}
  M_{fi} = \sum_{j=1,2} \int d^3 r_0 \: 
  \phi_{n}^{(-) \ast}(\vec{p}_n,\vec{r}_0) \, \langle \chi_n \mid 
  \langle \psi_{ab} \mid t_{0j} (\omega, \vec{r}_0 - \vec{r}_j )
  \mid \psi_d \rangle \mid \chi_p \rangle \, 
  \phi_p^{(+)} (\vec{p}_p,\vec{r}_0).
\end{equation}
Here $\vec{r}_j$ denotes the coordinates of the projectile $(j=0)$ 
and the target nucleons $(j=1,2)$, respectively, which are measured
relative to the center of mass of the target.
$\phi^{(+)}_p$ and $\phi^{(-)*}_n$ are the distorted wave functions
of projectile and ejectile in the initial and final channels. 
The spin--isospin part of the projectile (ejectile) wave function
is denoted as $\mid \chi_p \rangle$ ($\langle \chi_n \mid \, $). 
The initial and final states of the target are
$\mid \psi_d \rangle$ and $\langle \psi_{ab} \mid$, where
$ab$ refers to either the $NN$ or the $\Delta N$ system.
Note that $M_{fi}$ describes only the excitation process but not
the de--excitation, e.g.\ the free decay of the $\Delta$.
Therefore $M_{fi}$ has to be distinguished from the transition
amplitude $T_{fi}$ for the complete reaction.

The $t_{0j}$ in Eq.\ (\ref{eq3}) is the effective interaction
between the projectile nucleon $0$ and the target nucleon $j$.
A sum over the two target nucleons has to be performed.
The $t_{0j}$  is represented by the free nucleon--nucleon $t_{NN,NN}$ matrix
in the case of the nucleon excitation (Fig.\ \ref{fig1}a)
while it is approximated by the free $NN \rightarrow N\Delta$ 
transition operator $t_{NN,N\Delta}$ for the $\Delta$ excitation 
(Fig.\ \ref{fig1}b).
The specific form of these interactions is given in App.\ \ref{app:1}.
They provide a spin--longitudinal (pion--like) and a spin--transverse
($\rho$--meson--like) excitation component. The excitation strength
parameters are fitted to reproduce the experimental cross section 
and the spin observables for $(p,n)$ reactions on a nucleon target.

Since the effective $t_{0j}$ is adjusted to experiment, we
implicitly account for knockout exchange effects between the 
projectile and the target nucleons. Note, however,
that we neglect the projectile $\Delta$ excitation process
where the detected neutron comes from the decay of the $\Delta$.
For forward neutron angles, the cross section contribution of 
projectile excitation gives only a small correction to the 
dominant target excitation process. In Refs.\ \cite{jain93,jo96}, 
the projectile excitation is found to be suppressed by a factor of
$\approx 10$ in the $\Delta$ energy region.  

We now take advantage of the fact that at high incident energies 
and large momentum transfers both interactions turn out to be rather short 
ranged, i.e.\ very weakly dependent on the four--momentum transfer 
$(\omega,\vec{q}\, ) \equiv (E_p-E_n,\vec{p}_p-\vec{p}_n)$.
Therefore they can be well approximated by local operators in r--space of
essentially $\delta$--function form, that is
\begin{equation}
  \label{eq4}
  t_{0j}(\omega,\vec{r}_0-\vec{r}_j) =
  \frac{1}{(2\pi )^3} \int d^3 q\,' \: 
  \exp[i\vec{q}\;'\cdot(\vec{r}_0-\vec{r}_j)] \:
  t_{0j}(\omega,\vec{q}\,') \approx   
  t_{0j}(\omega,\vec{q}\,) \: \delta^3(\vec{r}_0-\vec{r}_j) .
\end{equation}
By inserting Eq.\ (\ref{eq4}) into Eq.\ (\ref{eq3}) we find for the 
transition amplitude
\begin{equation}
  \label{eq5}
  M_{fi} = \langle \psi_{ab} \mid \hat{\rho} \mid \psi_{d} \rangle,
\end{equation}
where the hadronic transition operator $\hat{\rho}$ is defined as
\begin{equation}
  \label{eq6}
  \hat{\rho}= \sum_{j=1,2} \langle \chi_n \mid
  t_{0j} (\omega,\vec{q}\,) \: X_{DW}(\vec{p}_p,\vec{p}_n,\vec{r}_j)
  \mid \chi_p \rangle ,
\end{equation}
with
\begin{equation}
  \label{eq7}
   X_{DW}(\vec{p}_p,\vec{p}_n,\vec{r}_j\,) =
  \phi_n^{\ast}(\vec{p}_n,\vec{r}_j\,) \phi_p (\vec{p}_{p},\vec{r}_j\,).
\end{equation}

The product wave function $X_{DW}$ includes the projectile 
and ejectile distortion effects in the reaction. 
In the eikonal approximation, Eq.\ (\ref{eq7}) is reduced to 
$X_{DW}(\vec{q}, \vec{r}_j) = N_{DW}(\vec{q}\,) 
\exp{(i\vec{q}\cdot\vec{r}_j\,)}$,
where $N_{DW} (\vec{q}\,)$ can be calculated from the distorting potential.
Thus we get from Eq.\ (\ref{eq6})
\begin{equation}
  \label{eq8}
  \hat{\rho} = N_{DW} (\vec{q}\,) \, \langle \chi_n \mid 
  t_{01}(\omega,\vec{q}\,) \exp{(i\vec{q}\cdot\vec{r}/2)} +
  t_{02}(\omega,\vec{q}\,) \exp{(-i\vec{q}\cdot\vec{r}/2)} 
  \mid \chi_p \rangle ,
\end{equation}
where $\vec{r} = \vec{r}_1 - \vec{r}_2$ is the relative coordinate between 
the two target nucleons.

Following Ref.\ \cite{udagawa94}, we can now introduce source functions 
for the excitation of either a $NN$ or a $\Delta N$ system, respectively, 
by defining
\begin{mathletters}
  \begin{eqnarray}
    \label{eq9a}
    \mid \rho_N (\vec{r} \,) \rangle &=& 
    N_{DW} (\vec{q}\,) \: \langle \chi_n \mid 
    t_{01}^{NN,NN} (\omega,\vec{q}\,) \exp{(i\vec{q}\cdot\vec{r}/2)}
    \mid \chi_p \rangle \mid \psi_d(\vec{r} \,) \rangle
    + ( 1 \leftrightarrow 2) ,\\
   \label{eq9b}
   \mid \rho_{\Delta} (\vec{r} \,) \rangle &=& 
    N_{DW} (\vec{q}\,) \: \langle \chi_n \mid 
    t_{01}^{NN,N\Delta} (\omega,\vec{q}\,) \exp{(i\vec{q}\cdot\vec{r}/2)}
    \mid \chi_p \rangle \mid \psi_d(\vec{r} \,) \rangle
    + ( 1 \leftrightarrow 2) .
  \end{eqnarray}
\end{mathletters}
The source functions represent the doorway states excited initially by
the external $(p,n)$ charge exchange field. We call them uncorrelated
since they do not include the ``final state'' interactions within the 
excited $NN$ or $\Delta N$ system. If the final state interactions are 
neglected, the second target nucleon is just a spectator which does not 
take part in the interaction with the projectile (``spectator approximation''). 
The full dynamical treatment of the $NN$ and $\Delta N$ systems 
requires the calculation of correlated continuum wave functions which 
we discuss in the next section.


\subsection{Inclusion of the $\Delta N$ interaction and calculation of the 
correlated wave function}

The inclusion of final state interactions in the description of the
$^2$H$(p,n)$ reaction leads to a coupled channel problem. In the energy
regime considered here, up to four particles appear in the final 
state because of the pion production. Therefore the full coupled
channel problem is complicated and can only be solved approximately.
In our model, we will treat the $\Delta$ as a quasi--particle with 
a given intrinsic decay width and mass.
The configuration space $\cal H$ is build up from $NN$, $\Delta N$,
and $\pi NN$ sectors, 
${\cal H} = {\cal H}_{NN} \oplus {\cal H}_{\Delta N}
\oplus {\cal H}_{\pi NN}$.
The corresponding projection operators are denoted as 
$P_N$, $P_\Delta$, and $P_Q$, respectively, and we will use for 
any operator $\Omega$ the obvious notation 
$\Omega_{\Delta N} = P_{\Delta} \Omega P_N$, 
$\Omega_{Q \Delta} = P_Q \Omega P_{\Delta}$, etc. 

In this formulation the source functions of Eqs.\ (\ref{eq9a},\ref{eq9b}) 
appear to be projections of a unique source function $\mid \rho \rangle$, 
i.e.
\begin{equation}
   P_N \mid \rho \rangle = \mid \rho_N \rangle, \quad 
   P_\Delta \mid \rho \rangle = \mid \rho_\Delta \rangle .
\end{equation}
Since we restrict ourselves to DWIA there is no source function 
for the $\pi NN$ sector, i.e. $P_Q \mid \rho \rangle = 0$.
We now define the correlated continuum wave function 
$\mid \psi \rangle$ as 
\begin{equation}
  \label{eq10}
  \mid \psi \rangle = \frac{1}{E-H+i \epsilon} \mid \rho \rangle 
  = G \mid \rho \rangle .
\end{equation}
The Hamiltonian $H=H^0+V$ contains the free energy $H^0$
of the system as well as the interaction V. The full propagator 
G is connected to the unperturbed Greensfunction $G^0 = (E-H^0)^{-1}$ 
by the equation 
\begin{equation}
  G = G^0 + G^0 V G .
  \label{eq11}
\end{equation}
The interaction V introduces the correlation effects into the wave 
function $\mid \psi \rangle$. We will demonstrate next how
to solve Eq.\ (\ref{eq10}) in a very efficient way
by using a few simplifying but physically reasonable approximations. 

First we make use of the fact that pion production in intermediate energy 
charge exchange reactions is dominated by the $\Delta$ resonance excitation. 
We assume that $V_{NQ} = V_{QN} = 0$, i.e. there is no direct coupling 
between the $NN$ and the $\pi NN$ sector. 
This means that in our model pion production proceeds exclusively via an 
intermediate $\Delta N$ system while the nucleon pole terms are neglected.
In the $\Delta$ excitation energy regime this approximation should be very 
good. 

The correlated $\Delta N$ wave function $\mid \psi_\Delta \rangle $
follows from Eq.\ (\ref{eq10}) by projecting on the $\Delta N$ channel,
i.e.\ \mbox{ $\mid \psi_\Delta \rangle = P_\Delta \mid \psi \rangle$ }. 
In the $\Delta$ energy region, $\mid \psi_\Delta \rangle$ can be well 
approximated by
\begin{equation}
  \label{eq12}
  \mid \psi_\Delta \rangle \approx G_{\Delta \Delta} 
  \mid \rho_\Delta \rangle .
\end{equation}
Here we neglected the contribution $G_{\Delta N} \mid \rho_N \rangle$
from the $NN$ source function which is substantially suppressed in the 
$\Delta$ energy region. 
The propagator $G_{\Delta \Delta}$ of the $\Delta N$ system is 
determined from Eq.\ (\ref{eq11}) which yields  
\begin{equation}
  \label{eq13}
  G_{\Delta\Delta} = G^0_{\Delta\Delta} 
                   + G^0_{\Delta\Delta} V_{\Delta\Delta} G_{\Delta\Delta} 
                   + G^0_{\Delta\Delta} V_{\Delta Q} G_{Q \Delta} 
                   + G^0_{\Delta\Delta} V_{\Delta N} G_{N \Delta} .
\end{equation}
In this equation, the first and second term on the r.h.s.\ represent 
the free propagator and the interaction term in the $\Delta N$ channel,
respectively.
The third term $G^0_{\Delta\Delta} V_{\Delta Q} G_{Q \Delta}$
reflects the emission and re--absorption of pions 
$(\Delta \rightarrow \pi N \rightarrow \Delta)$ which leads to the energy 
dependent decay width $\Gamma_\Delta (s_\Delta)$.  
However, since we treat the $\Delta$ as a quasi--particle
with an intrinsic width and the physical mass $M_\Delta =1232$ MeV, 
this ``self--dressing'' process is already effectively included in the free 
propagator $G^0_{\Delta \Delta}$. Therefore the third term of 
Eq.\ (\ref{eq13}) has to be dropped. In addition, we also drop the fourth 
term $G^0_{\Delta\Delta} V_{\Delta N} G_{N \Delta}$ which
generates $NN$ box contributions to the $\Delta N$ potential. We do
not expect those contributions to influence the $\Delta N$ interaction
significantly since it seems to be much more likely that the $\Delta$ is 
produced directly in the interaction with the projectile than later on
in the final state interaction of two outgoing nucleons. 
Hence we are left with
\begin{equation}
  \label{eq13a}
  G_{\Delta\Delta} \approx G^0_{\Delta\Delta} 
  + G^0_{\Delta\Delta} V_{\Delta\Delta} G_{\Delta\Delta} 
\end{equation}
for the $\Delta N$ propagator. This approximation has the major 
advantage that the $\Delta N$ channel effectively decouples from the 
$NN$ channel and therefore can be solved separately. 

If we project now Eq.\ (\ref{eq10}) to the possible final channels
$NN$ or $\pi NN$ and make use of Eq.\ (\ref{eq13a}), we find
\begin{mathletters}
  \label{eq14}
  \begin{eqnarray}
    \label{eq14a}
    P_N \mid \psi \rangle &=& G_{NN} \mid \rho_N \rangle  
      + G_{NN} V_{N \Delta} G_{\Delta \Delta} \mid \rho_\Delta \rangle , \\
    \label{eq14b}
    P_Q \mid \psi \rangle &=& G_{QQ} V_{Q \Delta} G_{\Delta \Delta} 
      \mid \rho_\Delta \rangle ,
  \end{eqnarray}
\end{mathletters}
with
\begin{mathletters}
  \begin{eqnarray}
    \label{eq15a}
    G_{NN} &=& G^0_{NN} + G^0_{NN} V_{NN} G_{NN} , \\ 
    \label{eq15b}
    G_{QQ} &=& G^0_{\pi} \otimes G_{NN} .
  \end{eqnarray}
\end{mathletters}
Transitions from the $\Delta N$ to the $NN$ or $\pi NN$ channel 
are accounted for by the corresponding transition potentials 
$V_{N \Delta}$ and $V_{Q \Delta}$ in Eq.\ (\ref{eq14}). They appear
only in first order.
The potential $V_{NN}$ in Eq.\ (\ref{eq15a}) describes the $NN$ final
state interaction which includes also $\Delta N$ box contributions. 
We adopt the Paris potential \cite{lacombe80,lacombe81} for $V_{NN}$. 
All effects of the $NN$ potential can be included in the $NN$ wavefunction 
$\langle \psi_{NN} \mid$ for the outgoing nucleons. 
Furthermore, we assume in Eq.\ (\ref{eq15b}) that the outgoing pion  
is not distorted from the remaining $NN$ system so that it can be 
described by a plane wave.

In the remainder of this section, we focus on the $\Delta N$ subsystem 
and the method for solving
\mbox{ $\mid \psi_{\Delta} \rangle = G_{\Delta \Delta} 
\mid \rho_\Delta \rangle $ }.
After having separated the c.m.\ motion, the full propagator for the relative 
$\Delta N$ wavefunction is given by
\begin{equation}
  \label{eq15c}
  G_{\Delta\Delta} = \frac{1}{ \epsilon_\Delta + \frac{i}{2} 
  \Gamma_\Delta (s_{\Delta}) - \hat{T}_{\Delta} - V_{\Delta \Delta} } , 
\end{equation}
with the excitation energy
\begin{equation}
  \epsilon_{\Delta} = \omega + M - M_{\Delta} 
  - \frac{(\vec{P}_{\text{cm}}) ^2}{2 (M+M_{\Delta})}
\end{equation}
and the relative kinetic energy
\begin{equation}
  \hat{T}_{\Delta} =  \frac{M + M_\Delta}{2 M M_\Delta} \: \vec{p}^{\, 2} . 
\end{equation}
In the spirit of the model, the propagator contains the 
energy dependent width $\Gamma_\Delta (s_\Delta)$ of the $\Delta$ 
resonance. The invariant mass $s_{\Delta}$ is fixed by  
conservation of four--momentum at the production vertex. 
For given projectile kinematics we obtain 
\begin{equation}
  s_{\Delta} = (M + \omega_{\text{lab}})^2 - \vec{q}^{\: 2}_{\text{lab}}. 
\end{equation}
Here $\omega$ is the energy transfer and $\vec{q}$ the momentum transfer
to the deuteron target. Note that the invariant $\Delta$ mass is calculated 
by assuming the target nucleon to be at rest in the laboratory frame
(frozen approximation).

In order to solve 
$\mid \psi_{\Delta} \rangle = G_{\Delta \Delta} \mid \rho_{\Delta} \rangle$, 
we transform this equation into an equivalent integral equation
\cite{udagawa94}
\begin{equation}
  \label{eq16}
  \mid \Lambda_{\Delta} \rangle = \mid \rho_{\Delta} \rangle  
  + V_{\Delta\Delta} G^0_{\Delta\Delta} \mid \Lambda_{\Delta} \rangle ,
\end{equation}
so that
\begin{equation}
  \label{eq17}
  \mid \psi_{\Delta} \rangle = G^0_{\Delta\Delta} 
  \mid \Lambda_{\Delta} \rangle .
\end{equation}

Eq.\ (\ref{eq16}) is now reduced to a set of coupled channel equations for 
radial wave functions in the following way. First we expand 
$\mid \rho_\Delta \rangle$ and $\mid \Lambda_\Delta \rangle$ 
in terms of partial waves, i.e.
\begin{equation}
  \label{eq18}
  \mid \rho_{\Delta} \rangle = 
  \sum_{SLJM_J} \frac{1}{r} \rho_{SLJM_J} (r) \mid (SL)JM_J \rangle
  \mid 11 \rangle ,
\end{equation}
\begin{equation}
  \label{eq18a}
  \mid \Lambda_{\Delta} \rangle = 
  \sum_{SLJM_J} \frac{1}{r} \lambda_{SLJM_J} (r) \mid (SL)JM_J \rangle
  \mid 11 \rangle ,
\end{equation}
where $S,L,J,M_J$ denote the spin, orbital and total angular momentum
quantum numbers of the $\Delta N$ system.
Due to the special isospin structure of the excitation process, the 
isospin channel is always $\mid T M_T \rangle = \mid 11 \rangle$.
After insertion of Eqs.\ (\ref{eq18},\ref{eq18a}) into Eq.\ (\ref{eq16})
and projection on $\langle (S'L') J' M_J' \mid$ we obtain 
\begin{equation}
  \label{eq19}
  \frac{1}{r} \, \lambda_n (r) = \frac{1}{r} \, \rho_n (r) + 
  \sum_{n'} \int dr' \, r'^{\, 2} \:
  V_{nn'} (r) \, G^0_{\Delta\Delta} (r,r\,') \, 
  \frac{1}{r\,'} \, \lambda_{n'} (r\,') ,
\end{equation}
where we used the index n as a shorthand notation for $\{ SLJM_J \}$.
Eq.\ (\ref{eq19}) may easily be written as a matrix equation.
The radial source functions $\rho_n (r)$ can be calculated in a 
straightforward manner and are given in App.\ \ref{app:2}. 
The matrix elements $V_{nn'} (r)$ of the $\Delta N$ potential, which is 
discussed in the next section, are given explicitly in App.\ \ref{app:3}.
Note that the operation of $G^0_{\Delta\Delta}$ onto $\lambda_{n'} (r')$ 
involves a radial integration besides the matrix multiplication.

The merit of solving for $\mid \Lambda_\Delta \rangle$ first lies in the 
fact that the corresponding radial functions $\lambda_n (r)$ are localized. 
This fact makes it possible to apply the Lanczos method
which is described in Ref.\ \cite{udagawa94}. It turns out that the Lanczos 
method allows us to solve Eq.\ (\ref{eq19}) in a very efficient way.
Once $\mid \Lambda_\Delta \rangle$ is known, it is easy to calculate 
$\mid \psi_{\Delta} \rangle$ from Eq.\ (\ref{eq17}) as well.


\section{The meson exchange model for the $\Delta N$ interaction}
\label{sec:vdn}

\begin{figure}
  \caption{Direct term (a) and exchange term (b) of the $\Delta N$ 
     potential $V_{\Delta \Delta}$. The mesons taken into account are
     the pion ($\pi$), the rho ($\rho$), the omega ($\omega$), and
     the sigma ($\sigma$).} 
  \label{fig2}
  \begin{center}
    \epsfig{file=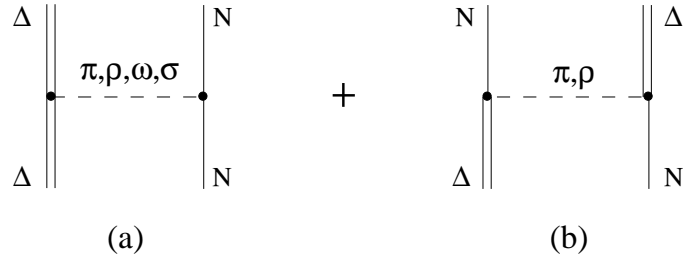, height=3.4cm}
  \end{center}
\end{figure}

Similar to the $NN$ interaction, the $\Delta N$ interaction can be 
constructed within a meson exchange model. 
In the present work we use $\pi$, $\rho$, 
$\omega$, and $\sigma$ exchange. While the $\omega$ and $\sigma$ mesons 
contribute only to the direct term shown in Fig.\ \ref{fig2}(a), 
the $\pi$ and $\rho$ mesons may also induce spin--isospin--flip transitions 
which lead to the exchange term of Fig. \ref{fig2}(b). 
We will discuss the latter term first and start from the
interaction Lagrangians ${\cal L}_{\pi N \Delta}$ and 
${\cal L}_{\rho N \Delta}$ given in Ref.\ \cite{machleidt87}.
In the non--relativistic reduction we obtain
\begin{equation}
  \label{eq20}
  V^{\pi}_{\text{ex}} (k) = \frac{f_{\pi N \Delta}^2}{m_{\pi}^2} 
  \: \vec{T_1} \cdot \vec{T_2}^{\dagger} \: 
  \frac{ (\vec{S_1} \cdot \hat k) (\vec{S_2}^{\dagger} \cdot \hat k) }
       { k_0^2 - \vec{k}^2 - m_{\pi}^2 + i \epsilon }
  + (1 \leftrightarrow 2) , 
\end{equation}
\begin{equation}
  \label{eq21}
  V^{\rho}_{\text{ex}} (k) = \frac{f_{\rho N \Delta}^2}{m_{\rho}^2} 
  \: \vec{T_1} \cdot \vec{T_2}^{\dagger} \: 
  \frac{(\vec{S_1} \times \hat k) \cdot (\vec{S_2}^{\dagger} \times \hat k)}
       {k_0^2 - \vec{k}^2 - m_{\rho}^2 + i \epsilon} 
  + (1 \leftrightarrow 2) .
\end{equation}
The corresponding direct interaction terms
follow from these potentials by replacing the transition matrices 
$\vec{S}$, $\vec{S}^{\dagger}$ ($\vec{T}, \vec{T}^{\dagger}$)
with the spin (isospin) 3/2 matrix $\vec{\Sigma}$ ($\vec{\Theta}$) and the
spin (isospin) 1/2 matrix $\vec{\sigma}$ ($\vec{\tau}$), respectively. 
This yields
\begin{equation}
  \label{eq23}
  V^{\pi}_{\text{dir}} (k) = 
  \frac{f_{\pi \Delta \Delta} f_{\pi NN}}{m_{\pi}^2} 
  \: \vec{\Theta_1} \cdot \vec{\tau_2} \:
  \frac{ (\vec{\Sigma_1} \cdot \hat k) (\vec{\sigma_2} \cdot \hat k) }
       { k_0^2 - \vec{k}^2 - m_{\pi}^2 + i \epsilon }
  + (1 \leftrightarrow 2) , 
\end{equation}
\begin{equation}
  \label{eq24}
  V^{\rho}_{\text{dir}} (k) = 
  \frac{f_{\rho \Delta \Delta} f_{\rho NN}}{m_{\rho}^2} 
  \: \vec{\Theta_1} \cdot \vec{\tau_2} \:
  \frac{ (\vec{\Sigma_1} \times \hat k) \cdot (\vec{\sigma_2} \times \hat k) }
       { k_0^2 - \vec{k}^2 - m_{\rho}^2 + i \epsilon } 
  + (1 \leftrightarrow 2) .
\end{equation}

For explicit calculation of these potentials we need to know the appropriate 
coupling constants. The $\pi NN$ and $\rho NN$ couplings
are well known experimentally from an analysis of $NN$ scattering data
\cite{machleidt87}. They may be related to the other couplings by  
\begin{mathletters}
  \begin{eqnarray}
    \label{eq25a}
    f_{\pi / \rho N \Delta} &=& 2 f_{\pi / \rho NN} , \\
    \label{eq25b}  
    f_{\pi / \rho \Delta \Delta} &=& \frac{1}{5} f_{\pi / \rho NN} . 
  \end{eqnarray}
\end{mathletters}
Eq.\ (\ref{eq25b}) follows from the static quark model \cite{brown75}.
For Eq.\ (\ref{eq25a}) we decided to use the widely accepted 
Chew--Low relation \cite{chew56} instead because the quark model prediction
for $f_{\pi N \Delta}$ is too small as compared to the experimental
value from the $\Delta$ decay.

Additional contributions to the direct part of the $\Delta N$ interaction
emerge from the $\omega$ and $\sigma$ exchange.
We assume the couplings $\omega \Delta\Delta$ and $\sigma \Delta \Delta$
to be the same as for $\omega NN$ and $\sigma NN$, respectively.   
As before, we choose the non--relativistic limit of the interaction which
follows from ${\cal L}_{\omega NN}$ and ${\cal L}_{\sigma NN}$
\cite{machleidt87}. The resulting potential is
\begin{equation}
  \label{eq26}
  V^{\sigma + \omega}_{\text{dir}} (k) =
    + g_{\omega \Delta \Delta} g_{\omega NN} \,
      \frac{1}{k_0^2 - \vec{k}^2 - m_{\omega}^2 + i \epsilon}
    - g_{\sigma \Delta \Delta} g_{\sigma NN} \, 
      \frac{1}{k_0^2 - \vec{k}^2 - m_{\sigma}^2 + i \epsilon} .
\end{equation}

The full potential for the $\Delta N$ interaction is the sum of
the different contributions from Eqs.\ (\ref{eq20} -- \ref{eq24}) and
(\ref{eq26}), hence
\begin{equation}
  V_{\Delta\Delta} (k) = V^{\pi+\rho}_{\text{ex}} (k) + 
  V^{\pi+\rho}_{\text{dir}} (k) + V^{\omega+\sigma}_{\text{dir}} (k) .
\end{equation}
Note that the $\pi$ ($\rho$) contribution covers the LO (TR) spin--isospin
part of the potential while the $\omega + \sigma$ contribution is 
spin--isospin independent. Furthermore, we point out that we use  
monopole form factors of the type
\begin{equation}
  F(k^2) = \left( \frac{\Lambda^2 - m^2}{\Lambda^2 - k_0^2 + \vec{k}^2} 
           \right)
\end{equation}
at all vertices. In the formulas given here, the form factors are 
always suppressed in order to simplify the notation.

Since the $\Delta N$ potential is required in r--space, a Fourier 
transformation of $V_{\Delta \Delta} (k)$ has to be performed.
This involves an integration over all possible three momentum transfers 
$\vec{k}$. On the other hand, the energy transfer $k_0$ is a free parameter
which has to be fixed within the kinematical allowed region. 
Because the $\Delta$ resonance is assumed to keep its 
invariant mass during the propagation, we choose $k_0 = 0$ in the direct 
interaction terms and $k_0 = \omega$ in the exchange terms.
This choice makes the potential energy--dependent, as it should be.

The $\Delta N \rightarrow NN$ transition potential $V_{N \Delta}$
is constructed in complete analogy from Eqs.\ (\ref{eq20}, \ref{eq21})
by replacing $\vec{S}^{\dagger}$ ($\vec{T}^{\dagger}$) with $\vec{\sigma}$
($\vec{\tau}$). Of course, only the $\pi$ and the $\rho$ meson contribute
to $V_{N \Delta}$. For the energy transfer in this case, 
we adopt the choice $k_0 = \omega / 2$ of Ref.\ \cite{maxwell80}. 
As far as the spin--longitudinal part of $V_{N \Delta}$ is concerned, 
we use an additional zero--range interaction 
\begin{equation}
  \label{eq27}
  V^{\delta}_{N \Delta} (k) = g_{N \Delta} \frac{f_{\pi N \Delta}
  f_{\pi NN}}{m_{\pi}^2} \: \vec{T_1} \cdot \vec{\tau_2} \:
  (\vec{S_1} \cdot \hat k) (\vec{\sigma_2}^{\dagger} \cdot \hat k) ,
\end{equation}
with the Landau--Migdal parameter $g_{N \Delta} = 1/3$ \cite{udagawa94}. 
This additional contribution cancels the non--physical zero--range part of 
the $\pi$ exchange potential. In principle one should introduce a corresponding
Landau--Migdal term for $V_{\Delta\Delta}$ with a strength parameter
$g_{\Delta\Delta}$ as well. However, due to the presence of not only 
exchange but also direct contributions to the potential, the zero--range 
parts cancel anyway and $g_{\Delta\Delta}$ turns out to have no influence 
on the results. Therefore we decided to do without and chose
$g_{\Delta\Delta} = 0$.

In Tab.\ \ref{tab:parameter}, an overview of all the meson parameters 
used for the $\Delta N$ potentials is given. We made sure 
that with these parameters, our model consistently reproduces experimental 
results for pion absorption on the deuteron, $\pi^+ + d \rightarrow 2p$
\cite{mosbacher95}.


\section{Decomposition of the inclusive cross section}
\label{sec:decomp}

\begin{figure}
  \caption{Reaction mechanisms of the different contributions to the
     inclusive cross section in our analysis. Only the lowest--order
     diagrams are shown. They represent: 
     (a) quasi--elastic scattering, (b) p--wave rescattering, 
     (c) coherent pion production, (d) quasifree $\Delta$ decay.}
  \label{fig3}
  \begin{center}
    \epsfig{file=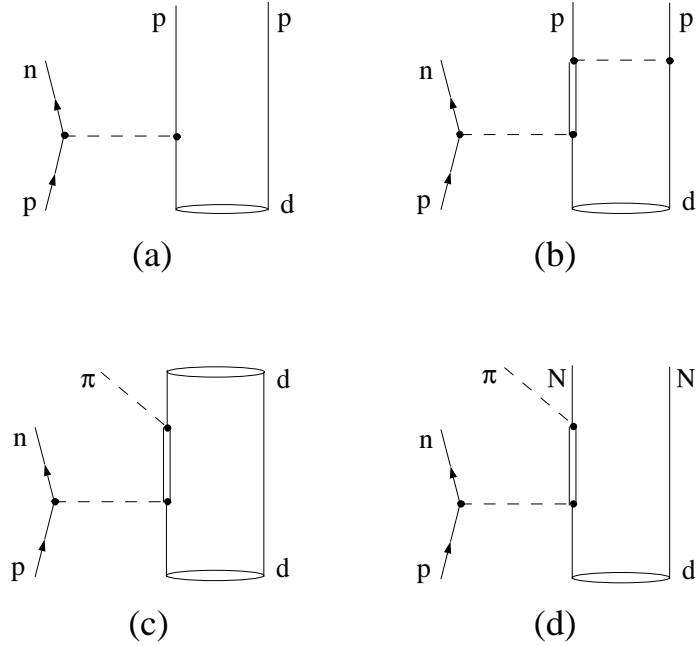, height=8.6cm}
  \end{center}
\end{figure}

We will now decompose the inclusive $^2$H$(p,n)$ cross section into partial
cross sections corresponding to different physical processes. These processes 
are schematically represented by the diagrams of Fig.\ \ref{fig3} (a)--(d). 
For each process, only the lowest--order diagram is shown.
We distinguish between quasi--elastic scattering (a), p--wave rescattering
(b), coherent pion production (c), and quasifree $\Delta$ decay (d).

Quasi--elastic scattering and p--wave rescattering both result in a 
two--proton ($2p$) final state on the target side. 
Their transition amplitudes interfere coherently with each other.
While the quasi--elastic scattering directly leads to the $2p$ system, 
the p--wave rescattering contribution involves an intermediate 
$\Delta N$ state and arises due to the $\Delta N \rightarrow NN$ 
transition potential $V_{N\Delta}$, see Fig.\ \ref{fig3}(a,b).
Since in both cases the deuteron is simply broken up into two protons, 
we will refer to this reaction channel as to the breakup channel (BU). 

At higher energy transfers, the $\pi NN$ final channel is the most 
important one. We remind the reader that we assume the pion production 
to proceed always through $\Delta$ resonance excitation. 
If the two nucleons in the final state form a bound deuteron state, 
we speak of coherent pion production (CP), see Fig.\ \ref{fig3}(c).
In this case, the outgoing pion is naturally a $\pi^+$. 
On the other hand, if the two outgoing nucleons are unbound, we speak of  
quasifree $\Delta$ decay (QF), see Fig.\ \ref{fig3}(d). 
There are two possible final isospin configurations for this process, 
namely $\pi^+np$ and $\pi^0 pp$. 

According to these different reaction mechanisms, the inclusive cross 
section of Eq.\ (\ref{eq2}) may now be split into its various components.
We express the squared transition matrix as the following sum, 
\begin{equation}
  \overline{\sum} \: \delta (\omega + E_d - E_f) \mid T_{fi} \mid^2 \, = 
  \frac{1}{6} \left( S_{\text{BU}} + S_{\text{CP}} + S_{\text{QF}} \right).
\end{equation}
Here the strength functions $S$ implicitly include the integration over 
free relative momenta as well as the summation over all spin configurations,
and the factor $\frac{1}{6}$ arises from the average over the initial spin
configurations.
 
Using Eq.\ (\ref{eq14a}), the strength function for the deuteron breakup 
channel BU is found to be
\begin{equation}
  S_{\text{BU}} = \sum_{S_i, S_f} 
    \int \frac{d^3 p}{(2 \pi)^3} \: \delta (\omega + E_d - E_{2p}) \: 
    \mid \langle \psi_{NN} \mid \rho_{N} \rangle 
    + \langle \psi_{NN} \mid V_{N \Delta} \mid \psi_{\Delta} \rangle \mid^2 ,
\end{equation}
The first matrix element in the sum of the r.h.s.\ describes the quasi--elastic
scattering, and the second matrix element describes the p--wave rescattering.
$\vec{p}$ denotes the relative momentum of the two outgoing protons.  
Their final state interaction $V_{NN}$ is included in $\langle \psi_{NN} \mid$ 
while the correlated source function $\mid \psi_{\Delta} \rangle$ 
accounts for the $\Delta N$ interaction $V_{\Delta \Delta}$.

In order to calculate the matrix element for coherent pion production, we
define the operator
\begin{equation}
  \label{eq30}
  \hat{f}_{\pi} = 
  e^{-i \vec{q}_{\pi} \cdot \vec{r} /2} \: \frac{f_{\pi N \Delta}}{m_{\pi}} \:
  T_{1 \nu} \, \vec{S_1} \cdot \vec{\kappa}_{\pi} + (1 \leftrightarrow 2) ,
\end{equation}
where $\vec{\kappa}_{\pi}$ is the relative pion--nucleon momentum
in the $\Delta$ rest frame. The operator 
describes the decay of the $\Delta$ into a real pion and a nucleon and 
contains also the plane wave of the outgoing pion. Applying $\hat{f}_\pi$
to the $\Delta N$ wave function and projecting onto the final deuteron state 
yields
\begin{equation}
  \label{eq31}
  S_{\text{CP}} = \sum_{S_i, S_f} 
  \int \frac{d^3 q_{\pi}}{(2 \pi)^3 \, 2E_{\pi}} 
  \: \delta (\omega + E_d - E_{\pi} - E_d' ) \: 
  \mid \langle \psi_{d} \mid \hat{f}_{\pi} \mid \psi_{\Delta} \rangle \mid^2
\end{equation}
for the CP strength function. Eq.\ (\ref{eq31}) may also be rewritten as  
\begin{equation}
  \label{eq31a}
  S_{\text{CP}} = \sum_{S_i} \:
  \langle \psi_\Delta \mid \frac{1}{\pi} \text{Im}(V_{\Delta \Delta})
  \mid \psi_\Delta \rangle ,
\end{equation}
where we used the fact that $E_d \approx E_d'$ holds in the
Breit frame. Hence the coherent pion production amplitude is directly 
connected to the imaginary part of the $\Delta N$ interaction. We remark
that this is only true if $\text{Im}(V_{\Delta \Delta})$ comes exclusively 
from the $\pi$ exchange interaction, as it is the case in the 
$\Delta$ energy region. 

From Eq.\ (\ref{eq31}) we may easily calculate the angular distribution 
of the coherent pions, namely the triple differential cross
section $d^3 \sigma_{CP} / dE_n d\Omega_n d\Omega_{\pi}$, by omitting the 
integration over the pion momentum. The result is
\begin{equation}
  \label{eq32}
  \left( \frac{d^3 \sigma_{CP}}{dE_n d\Omega_n d\Omega_{\pi}} 
  \right)^{\text{lab}} = 
  \frac{M_p M_n}{(2 \pi)^5} \, \frac{p_n^{\text{lab}}}{p_p^{\text{lab}}}
  \frac{E_d}{M_d} \, \frac{q^2_{\pi}}{2 E_{\pi}} \, \frac{dq_{\pi}}{dE_f} \,
  \mid \langle \psi_{d} \mid \hat{f}_{\pi} \mid \psi_{\Delta} \rangle \mid^2 
\end{equation}
with the same notation for the phase space factor as in Eq.\ (\ref{eq2}).

In analogy to Eq.\ (\ref{eq31a}), the pion production from quasifree 
$\Delta$ decay is ascribed to the imaginary part of the $\Delta$ 
``self--dressing'', i.e.\ to the intrinsic $\Delta$ decay width. 
As a consequence, the QF strength function is given as
\begin{equation}
  \label{eq34}
  S_{\text{QF}} = \sum_{S_i} \:
  \langle \psi_\Delta \mid \frac{1}{\pi} \frac{\Gamma_\Delta (s_{\Delta})}{2}
  \mid \psi_\Delta \rangle .
\end{equation}
The differential QF cross section could also be calculated from the 
matrix element 
\mbox{$\mid \langle \psi_{NN} \mid \hat{f}_\pi 
\mid \psi_\Delta \rangle \mid^2$}.
Eq.\ (\ref{eq34}) would then be recovered after integration over 
the pion momentum, as in Eq.\ (\ref{eq31}), and over the relative 
$NN$ momentum as well.

To summarize we have decomposed the inclusive $^2$H$(p,n)$ cross section 
into three parts: the quasifree $\Delta$ decay (QF), the coherent
pion production (CP) and the two--nucleon breakup of the deuteron 
(BU = quasi--elastic plus p--wave rescattering). 
Of course there are other possible contributions to the inclusive
cross section, but in the energy region under consideration,
they should be less important than the ones mentioned and are
therefore neglected in the present paper.


\section{Results and Discussion}


\subsection{Input parameters}

With the formalism described in the previous sections, we have calculated
energy spectra at various scattering angles $\theta_n$ for the $^2$H$(p,n)$ 
reaction at $T_p = 790$ MeV. 
The deuteron wave function and the two--nucleon wave 
function were generated from the Paris potential \cite{lacombe80,lacombe81}.
A constant distorsion factor of $N_{DW}^2 = 0.9$ was applied
\cite{chen93,jain93}. 
For the $\Delta N$ potential we used coupling constants and cutoff 
parameters as given in Table \ref{tab:parameter}.    
The free $\Delta$ mass is $M_{\Delta} = 1232$ MeV, and the free decay width 
$\Gamma_\Delta (s_\Delta)$ was parameterized in the usual form 
\cite{esbensen85}. 

\begin{figure}
  \caption{Differential cross section for the reaction 
     $^1$H$(p,n)\Delta^{++}$ at $T_p = 790$ MeV and at
     scattering angles $\theta_n = 0^o$,$ 7.5^o$, and $15^o$, respectively. 
     The theoretical calculation uses the parameterization of 
     $t_{NN,N \Delta}$ as given in App.\ \protect \ref{app:1}. 
     Experimental data are from Ref.\ \protect \cite{prout}.}
  \label{fig15}
  \begin{center}
    \epsfig{file=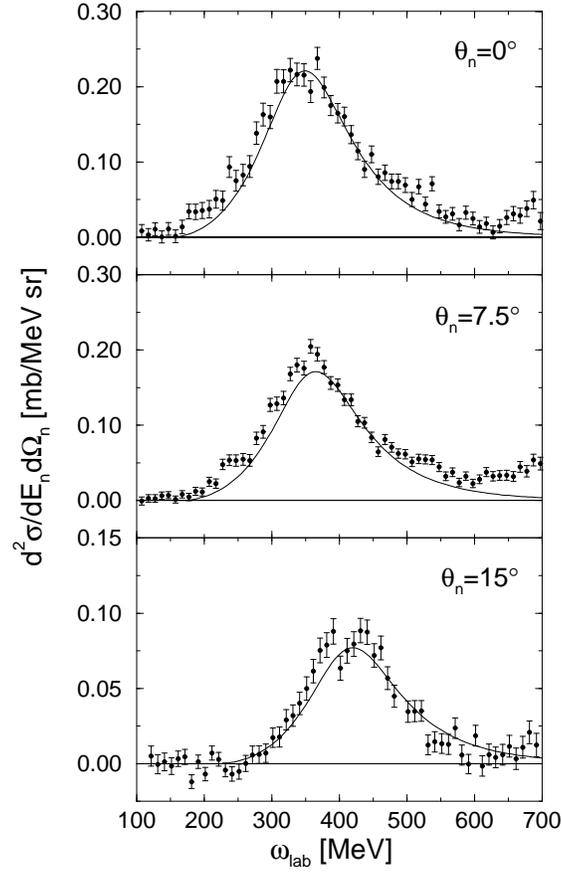, height=11.6cm}
  \end{center}
\end{figure}

The strength parameters in the effective parameterizations of $t_{NN,NN}$
and $t_{NN,N\Delta}$ (cp.\ Sec.\ \ref{sec:dwia}, App.\ \ref{app:1}) have been 
fitted to reproduce experimental results for the nucleon target. 
In order to demonstrate the quality of this fit, we present in 
Fig.\ \ref{fig15}
results for the $^1$H$(p,n)\Delta^{++}$ cross section at $T_p=790$ MeV.
The theoretical calculations are compared with the experimental data 
\cite{prout}
at various scattering angles $\theta_n$ of the outgoing neutron.
Both shape and magnitude are reproduced very well. We remark that also the 
LO/TR ratio of 1/2 used in the parameterization of $t_{NN,N\Delta}$ has been 
observed experimentally \cite{glass83,ellegaard89}.

\subsection{Inclusive spectra of the $^2$H$(p,n)$ reaction}

\begin{figure}
  \caption{Zero degree neutron spectrum for the reaction $^2$H$(p,n)$ at
     $T_p = 790$ MeV. The theoretical calculation (solid line) includes
     contributions from quasifree $\Delta$ decay (QF, dashed), coherent
     pion production (CP, dotted), and two--nucleon breakup of the 
     deuteron (BU, dashed--dotted).
     Experimental data are from Ref.\ \protect \cite{prout}.}
  \label{fig4}
  \begin{center}
    \epsfig{file=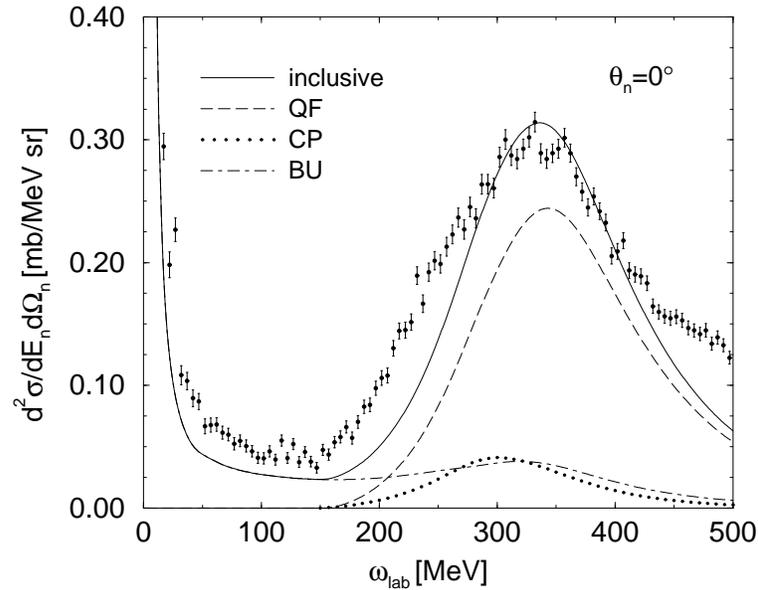, height=8.0cm}
  \end{center}
\end{figure}

In Fig.\ \ref{fig4} we show the calculated inclusive cross section 
for the $\theta_n = 0^o$ spectrum of the $^2$H$(p,n)$ reaction in comparison
with the experimental data \cite{prout}. The different contributions to the
inclusive cross section are also shown separately. The possible reaction
channels are the quasifree $\Delta$ decay (QF), the coherent pion production
(CP), and the deuteron breakup (BU) due to quasi--elastic scattering
and p--wave rescattering.

At energies $\omega_{\text{lab}} \leq 50$, the spectrum exhibits 
a prominent peak which arises dominantly from quasi--elastic scattering.
The small width of the peak reflects the Fermi motion of the nucleons 
in the deuteron target. The peak is described well by our calculations.

In the energy region of the $\Delta$ resonance peak position
at $\omega_{\text{lab}} = 340$ MeV, the theoretical result is also in
good agreement with the data. The cross section in this region is mainly 
due to the quasifree decay of the $\Delta$. The cross section contribution
due to coherent pion production is by a factor of $\approx 5$ smaller
as compared to the QF contribution. 

At high energy transfers $\omega_{\text{lab}} \geq 450$ MeV,
our calculation underestimates the experimental cross section.
The missing cross section can be ascribed to the 
excitation of higher nucleon resonances, e.g. $N^*(1440)$. 
Those configurations are not included in our model space, 
and therefore the underestimate of the cross section is not surprising. 

\begin{figure}
  \caption{Full calculation with inclusion of baryonic exchange currents
     (solid line) in comparison with the ``spectator approximation'' result 
     where $V_{\Delta\Delta} = V_{N\Delta} = 0$ (dashed line).
     Experimental data are from Ref.\ \protect \cite{prout}.}
  \label{fig5}
  \begin{center}
    \epsfig{file=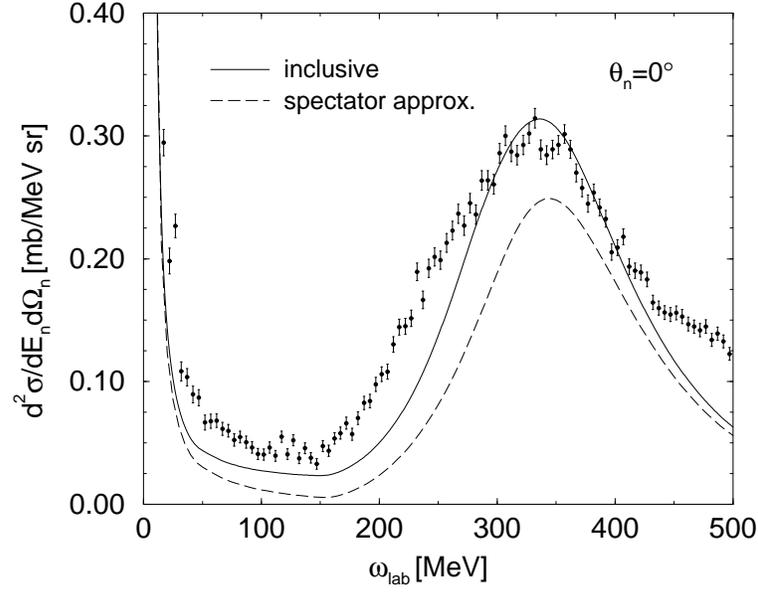, height=8.0cm}
  \end{center}
\end{figure}

The theoretical calculations also underestimate the data in the so called
dip--region between the quasi--elastic peak and the $\Delta$ peak,
i.e.\ in the energy region 50 MeV $\leq \omega_{\text{lab}} \leq 250$ MeV.
The experimental cross section here is believed to result
mainly from two--body exchange currents in the target. 
Only the baryonic exchange currents connected with the $\Delta$ are 
accounted for in our calculations. They are included in the $\Delta N$ 
interaction $V_{\Delta\Delta}$ and in the transition potential
$V_{N\Delta}$ which gives rise to the p--wave rescattering
contribution. In Fig.\ \ref{fig5}, our full model calculation is compared
with a calculation obtained within the ``spectator approximation''.
The latter one corresponds to the case $V_{\Delta\Delta}=V_{N\Delta}=0$. 
It can be seen that the inclusion of the exchange currents significantly
improves the theoretical description of the experimental data. Nevertheless,
we still underestimate the cross section in the dip--region by a factor 
of about 1.5. This may be due to the neglect of purely mesonic 
exchange currents, e.g. the s--wave rescattering of pions.
At the low energy side of the $\Delta$ peak, also relativistic corrections
in the dynamical treatment of the $\Delta N$ system could be of some 
importance.

\begin{figure}
  \caption{Separation of the inclusive cross section into spin--longitudinal
     (LO, $\sim \vec{\sigma} \cdot \hat{q}$) and spin--transverse
     components (TR, $\sim \vec{\sigma} \times \hat{q}$).
     Experimental data are from Ref.\ \protect \cite{prout94}.}
  \label{fig6}
  \begin{center}
    \epsfig{file=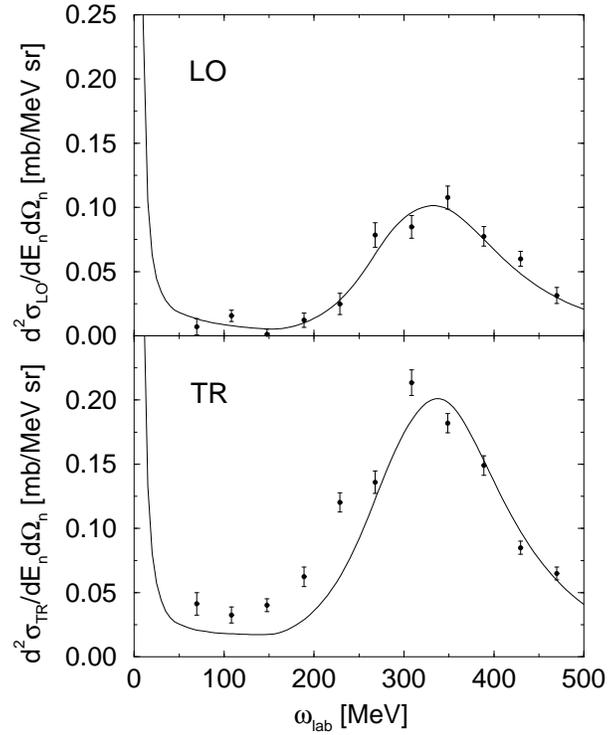, height=10.0cm}
  \end{center}
\end{figure}

In the following we shall demonstrate that almost 
all of the cross section in the dip--region is spin--transverse (TR) while 
the spin--longitudinal (LO) cross section is relatively small.
This statement can be tested by measuring
spin observables that allow for a separation of the inclusive
cross section into its LO and TR components \cite{prout94}.
This separation is shown in Fig.\ \ref{fig6} together with
the theoretical results.
The measured LO cross section is reproduced very well by our calculation.
On the other hand, the measured TR cross section is both
shifted and enhanced with respect to the calculation. 
This apparent shift clearly indicates that our calculations produce 
not enough TR cross section in the dip--region. A similar effect is 
observed in inelastic electron scattering off nuclei \cite{connell87,boffi93}.
In the $(e,e')$ reactions, the virtual photon also probes the TR
response of the target. In scattering off nuclei, the experimental $(e,e')$ 
cross section in the dip--region is considerably larger than expected.
The additional cross section cannot be explained by one--body currents
and has to be ascribed to two--body currents \cite{vanorden81}. 
Such rescattering effects have also proven to be quite important 
for coherent pion photoproduction on the deuteron \cite{wilhelm96}.
In hadronic interactions, the $\rho$ meson should take over the
role of the photon. In a recent paper of C.-Y. Lee \cite{lee97}, 
the author found indeed a large contribution to the $^2$H$(p,n)$ inclusive 
cross section from the exchange of a $\rho$ meson which couples to a
``pion--in--flight'' in the deuteron. We could not reproduce this
result. Since there are many other possible meson exchange 
currents which have not been considered for the $^2$H$(p,n)$ reaction
so far, the dip--region puzzle remains a problem to be solved.

\begin{figure}
  \caption{Neutron spectra for the reaction $^2$H$(p,n)$ at $T_p = 790$ MeV
    at scattering angles of $\theta_n = 7.5^o$ and $\theta_n = 15^o$,
    respectively.
    Experimental data are from Ref.\ \protect \cite{prout}.}
  \label{fig7}
  \begin{center}
    \epsfig{file=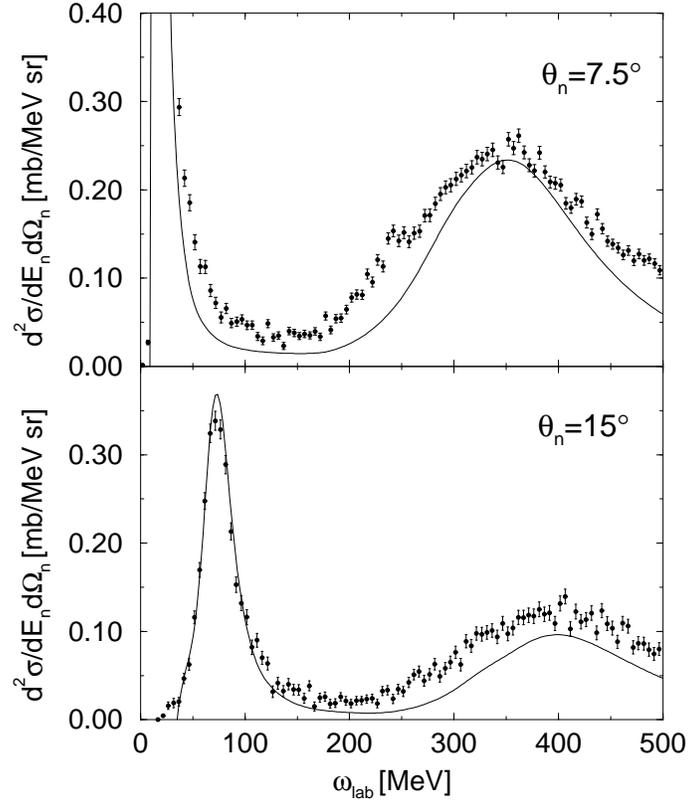, height=11.0cm}
  \end{center}
\end{figure}

The theoretical description of the experimental $^2$H$(p,n)$ spectra
at higher scattering angles is of the same quality as for 
zero degree scattering. This can be seen from Fig.\ \ref{fig7}
where we compare the theoretical $^2$H$(p,n)$ cross section with
data at $\theta_n = 7.5^o$ and $\theta_n = 15^o$, respectively.
At the low energy side of the $\Delta$,
the theoretical curves show again a characteristic underestimate 
of the data. Apart of this effect both the shape and the magnitude 
of the spectra are described well.
Our calculations correctly reproduce the change of  
the quasi--elastic and the $\Delta$ peak position with the 
scattering angle.


\subsection{Influence of the $\Delta N$ interaction on the exclusive spectra}

We now discuss the influence of the $\Delta N$ interaction on the three
partial contributions (QF, CP, BU) to the cross section.
As we will demonstrate, the coherent pion production is most sensitive to 
$V_{\Delta\Delta}$. This sensitivity is a combined effect of the attractive 
pion exchange, the spin--longitudinal $\pi N \Delta$ coupling and the 
structure of the deuteron wave function. In the following discussion,
we will focus on the exclusive $^2$H$(p,n\pi^+)^2$H spectra first.

\begin{figure}
  \caption{Effects on coherent pion production $^2$H$(p,n\pi^+)^2$H 
    resulting from the direct (dir) and the exchange (ex) contribution of the 
    $\pi$ meson to $V_{\Delta\Delta}$ (solid lines). The line labeled 
    (dir+ex) shows the sum of direct and exchange contributions
    of the $\pi$. Note the apparent shift as compared to the dashed 
    line which represents the spectator approximation $V_{\Delta\Delta}=0$.}
  \label{fig8}
  \begin{center}
    \epsfig{file=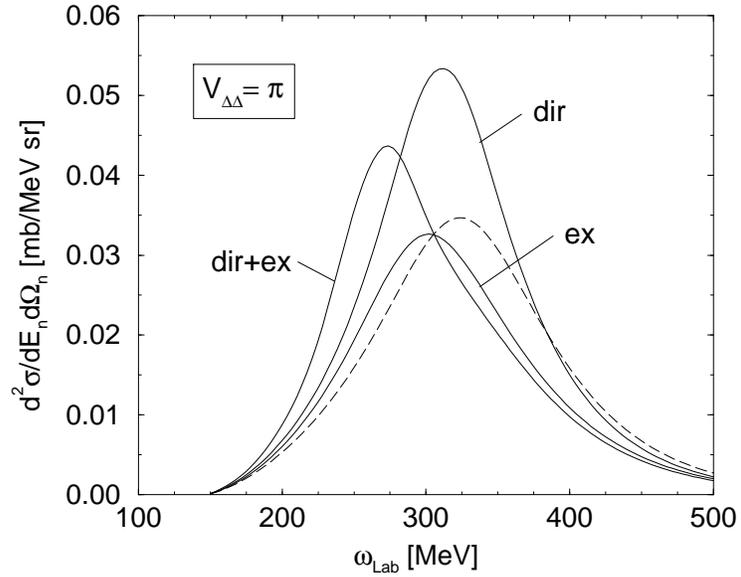, height=7.8cm}
  \end{center}
\end{figure}

In Fig.\ \ref{fig8}, theoretical results for the exclusive  
$^2$H$(p,n\pi^+)^2$H cross section are shown. The calculation without 
inclusion of the $\Delta N$ interaction 
(i.e. $V_{\Delta\Delta}=0$) is compared 
with a calculation where only the $\pi$ meson contribution to 
$V_{\Delta\Delta}$ was taken into account. The effects of the direct and 
the exchange part of the $\pi$ mediated interaction are also examined
separately. Obviously both parts are of equal importance which may be
surprising at first regard. It is true that because of
the smallness of the $\pi\Delta\Delta$ coupling constant,
the direct part should be suppressed by a factor of $\approx 20$ as 
compared to the exchange part,
but in the $T=1$ channel of the $\Delta N$ system which is relevant 
here, the spin--isospin matrix elements turn out to be larger by 
approximately the same amount. Therefore it is essential to include
both the direct and the exchange contribution in the $\Delta N$ interaction.
In the case of the $\pi$ exchange, the interference effect of both
contributions leads to a very strong attraction between the $\Delta$
and the second target nucleon. The result is a shift of the 
cross section peak position downwards in energy (by $\approx 60$ MeV,
cp.\ Fig.\ \ref{fig8}). 
Most of the attraction is caused by the tensor part of the $\pi$ whereas
the central part is less important for the peak position but
influences the overall magnitude of the result. 

\begin{figure}
  \caption{Effects on coherent pion production $^2$H$(p,n\pi^+)^2$H 
    resulting from the $\pi$, $\rho$, $\omega$, and $\sigma$ meson
    contributions to $V_{\Delta\Delta}$ (solid lines). 
    The line labeled ($\pi+\rho+\omega+\sigma$) represents the full model
    calculation. The dashed line shows the spectator approximation 
    $V_{\Delta\Delta}=0$.}
  \label{fig9}
  \begin{center}
    \epsfig{file=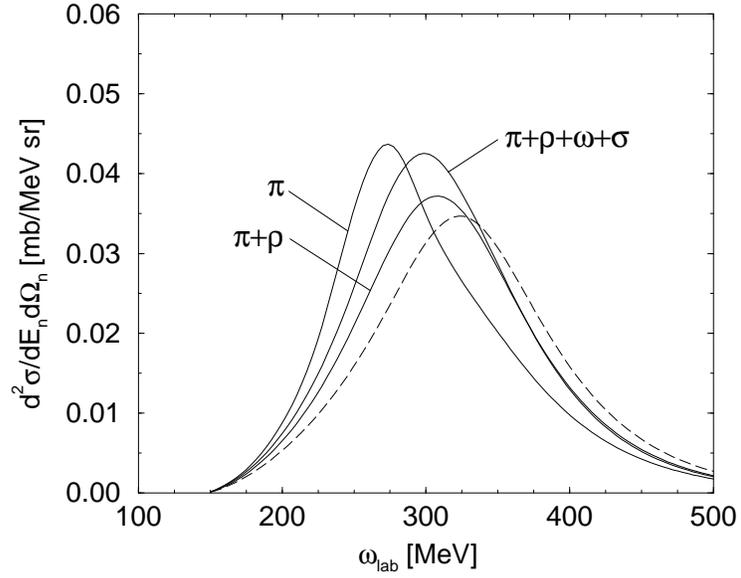, height=7.8cm}
  \end{center}
\end{figure}

By including not only the $\pi$, but also 
the $\rho$, $\omega$, and $\sigma$ mesons in $V_{\Delta\Delta}$, 
we obtain the results presented in Fig.\ \ref{fig9}. 
It can be seen that both the peak
position and the magnitude of the exclusive $^2$H$(p,n\pi^+)^2$H cross
section depend quite sensitively on the specific form of the
$\Delta N$ potential. 
In comparison to the result with only $\pi$ exchange, the 
inclusion of $\pi$ and $\rho$ in the $\Delta N$ interaction
leads to a less attractive potential and hence to a smaller
shift of the peak position. The reason for this behavior is
the partial cancelation of the $\pi$ and $\rho$ tensor forces
which have opposite sign.
For the direct part of the interaction, the $\omega$ and the $\sigma$
meson cause an additional short range repulsion and a medium
range attraction, respectively, which leads to an additional enhancement 
of the cross section. 
The final result for the $^2$H$(p,n\pi^+)^2$H spectrum (i.e. the result with 
inclusion of $\pi+\rho+\omega+\sigma$ in $V_{\Delta \Delta}$) is quite 
different from the spectator approximation (i.e. the result with 
$V_{\Delta\Delta}=0$).
Most remarkable is the shift of the peak position downwards in
energy by about 30 MeV. As discussed above, the main reason for this shift 
is the attractive pion exchange, i.e.\ the attractive LO part of the
$\Delta N$ potential. The amount of the peak shift is directly related
to the interaction strength of $V_{\Delta\Delta}$.  

\begin{figure}
  \caption{Effects of $V_{\Delta\Delta}$ on the exclusive 
     cross section for coherent pion production (CP),
     quasifree $\Delta$ decay (QF), and deuteron breakup (BU). 
     Calculations with (solid lines) and without $V_{\Delta\Delta}$ 
     (dashed lines).
     Note the different scalings of the vertical axis.}
  \label{fig10}
  \begin{center}
    \epsfig{file=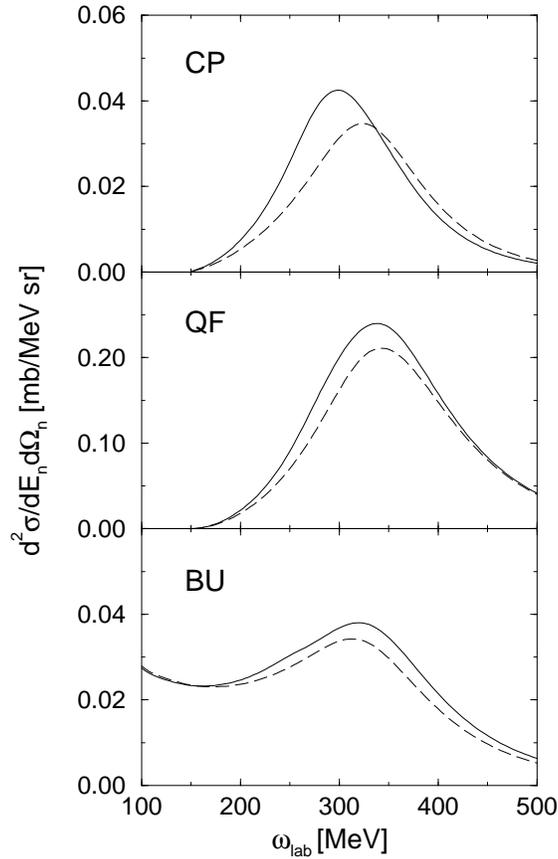, height=11.8cm}
  \end{center}
\end{figure}

This result attracts further interest because there is no comparable
peak shift due to $V_{\Delta\Delta}$ for the quasifree $\Delta$ decay
cross section and also not for the deuteron breakup. Fig.\ \ref{fig10} 
shows the results with and without inclusion of $V_{\Delta\Delta}$
for the coherent pion production (CP), the quasifree $\Delta$ decay
(QF) and the deuteron breakup (BU). In both the QF and the BU channel, the 
cross section is slightly enlarged and broadened due to the $\Delta N$ 
interaction, but there is no significant shift of the peak position. 

\begin{figure}
  \caption{Decomposition of the exclusive cross sections into 
     contributions from unnatural parity states (UP) and 
     natural parity states (NP). 
     Calculations with (solid lines) and without $V_{\Delta\Delta}$ 
     (dashed lines).
     Note the different scalings of the vertical axis.}
  \label{fig11}
  \begin{center}
    \epsfig{file=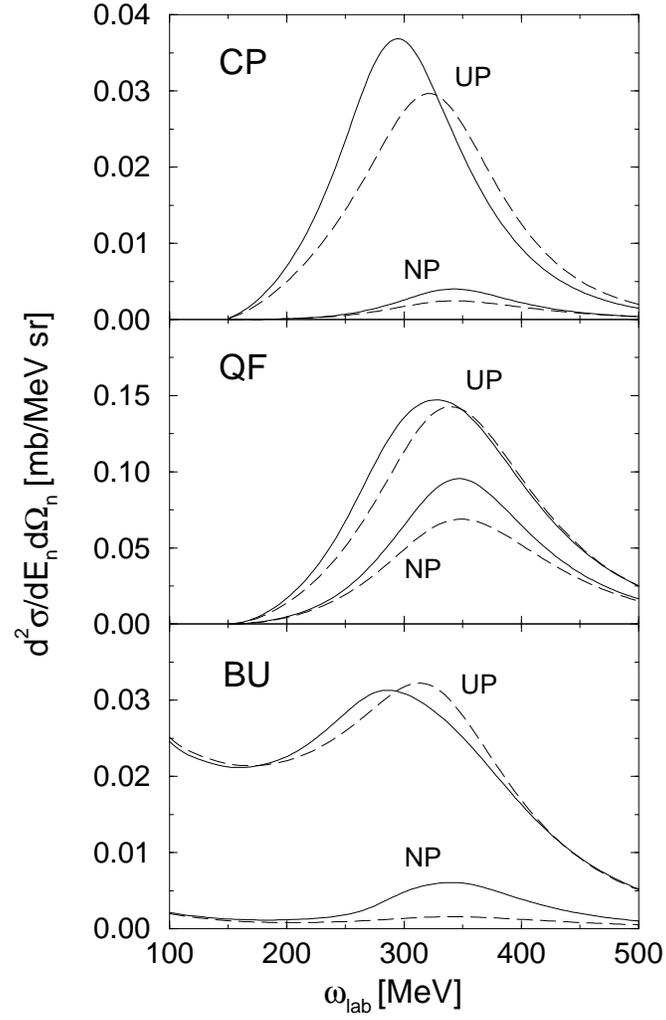, height=14.0cm}
  \end{center}
\end{figure}

In order to understand this different behavior, we perform a multipole
decomposition of the partial cross sections and split them into two 
contributions corresponding to unnatural parity (UP) states 
($J^P = 0^-, 1^+, 2^-, \ldots$) of the $\Delta$N system 
and to natural parity (NP) states ($J^P = 1^-, 2^+, \ldots$),
respectively. For the three reaction channels under 
consideration, this decomposition is shown in Fig.\ \ref{fig11}.
For each case, the full calculation is compared with the spectator 
approximation. One recognizes that the cross section contributions
of the UP states are always lowered in excitation energy
by the $\Delta N$ interaction while the NP states are not. 
This is explained by the fact that there is a strong
coupling of the pion to the UP (pion--like) states but just a weak 
coupling to the NP states. Therefore only the UP states are substantially 
influenced by the attraction of the LO part of $V_{\Delta \Delta}$.
For the coherent pion production, the spectrum
is clearly dominated by the UP states. This is an effect of the LO  
spin structure of the de--excitation process and of the 
deuteron wave function (which selects spin $S=1$ and orbital momentum 
$L=0,2$) in the final state.

For the quasifree $\Delta$ decay, the final deuteron state is replaced 
by a final unbound $NN$ state. Since the two outgoing nucleons are 
free, we have no particular spin selection rule in the de--excitation 
process for this case and thus all partial waves can contribute. 
As can be seen from Fig.\ \ref{fig11}, the NP states 
become more important for the QF decay, and the overall energy shift 
more or less disappears. 
For the $2p$ breakup, the $\Delta N \rightarrow NN$ transition 
potential $V_{N\Delta}$ provides not only a LO but also a TR component. 
The energy shift of the UP states is compensated by a relative large 
enhancement of the NP states which results from TR excitation. Moreover, 
the p--wave rescattering interferes with the quasi--elastic scattering
which has no intermediate $\Delta N$ configuration, hence the 
effects of $V_{\Delta\Delta}$ are partially smeared out.

\begin{figure}
  \caption{Multipole decomposition of the zero--degree spectrum
     for the exclusive $^2$H$(p,n\pi^+)^2$H reaction.
     Calculations with (solid lines) and without $V_{\Delta\Delta}$ 
     (dashed lines) for three characteristical partial waves of the 
     $\Delta N$ system are shown.
     Note the different scalings of the vertical axis.}
  \label{fig12}
  \begin{center}
    \epsfig{file=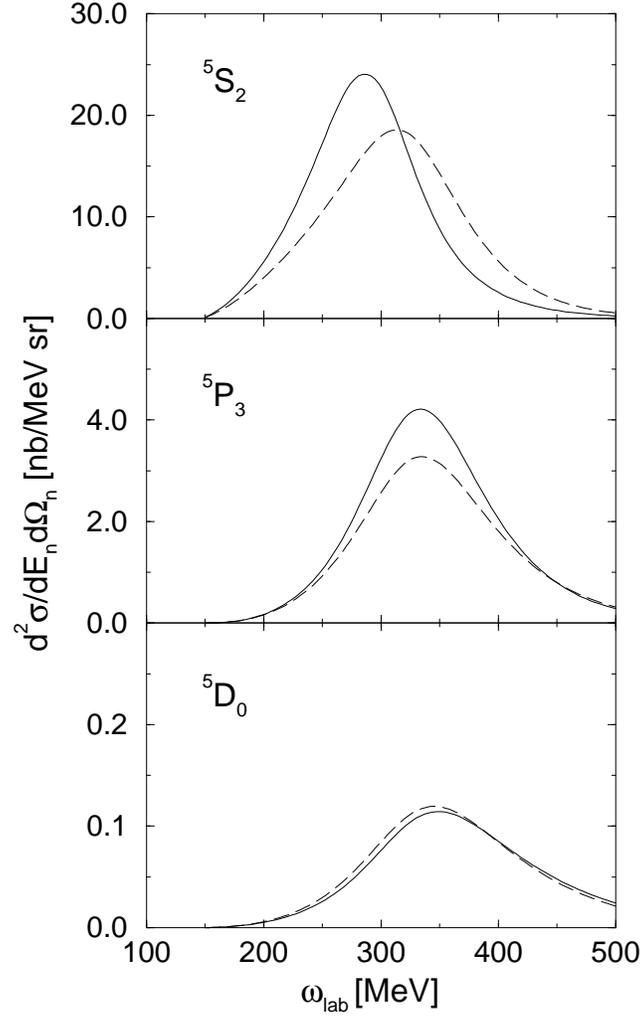, height=14.0cm}
  \end{center}
\end{figure}

To complete the discussion about the influence of the $\Delta N$
interaction, we show in Fig.\ \ref{fig12} the contributions 
of three characteristical $\Delta N$ partial waves to
the exclusive $^2$H$(p,n\pi^+)^2$H cross section. 
All selected partial waves have unnatural parity and therefore
couple strongly to the pion. The spectrum is
absolutely dominated by the $^5S_2$ partial wave of the 
$\Delta N$ system. Because the $\Delta$ and the nucleon have
relative angular momentum zero in this case, there is no
centrifugal barrier in the potential and the attraction 
becomes largest. That accounts for the strong energy shift
in this partial wave. The partial waves with higher angular
momentum $L \ge 1$ are subjected to a less attractive potential
due to the centrifugal barrier and therefore do not exhibit 
such a lowering in the excitation energy. As can be seen
from Fig.\ \ref{fig12}, the $^5P_3$ partial wave is only
enhanced but not shifted due to $V_{\Delta\Delta}$. For
the $^5D_0$ partial wave, the potential has a minor effect and 
even gets repulsive. 
Compared to $^5S_2$, the higher partial waves
are by far less important which results in the surviving
of the downwards energy shift in the full spectrum.

We conclude that neither the QF nor the BU process shows the same
sensitivity to the LO channel as the CP process.  
Therefore only the exclusive $^2$H$(p,n\pi^+)^2$H spectrum clearly exhibits 
a lowering of the $\Delta$ excitation energy. This fact makes the 
coherent pion production most suitable to examine the effects of the 
$\Delta N$ interaction.


\subsection{Exclusive differential cross section 
for coherent pion production}

\begin{figure}
  \caption{Triple differential cross section for the $^2$H$(p,n\pi^+)^2$H 
     reaction at $\omega_{\text{lab}} = 300$ MeV and $\theta_n = 0^o$.
     The cross section is shown as a function of $\theta_{\pi}$ which
     is the angle between the outgoing $\pi^+$ and the momentum transfer 
     $\vec{q}$ on the deuteron target.}
  \label{fig13}
  \begin{center}
    \epsfig{file=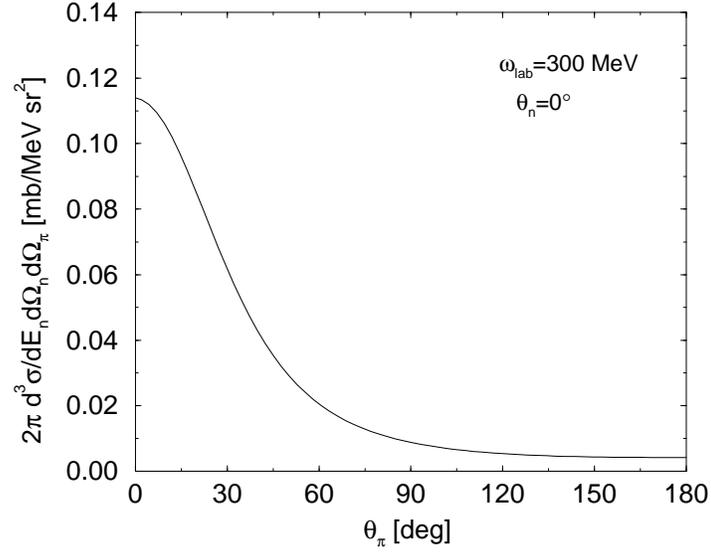, height=8.2cm}
  \end{center}
\end{figure}

In this section, we present our results for the angular distribution 
of the coherently produced pions. In Fig.\ \ref{fig13}, the
triple differential cross section $d^3 \sigma / dE_n d\Omega_n d\Omega_\pi$
of the exclusive $^2$H$(p,n\pi^+)^2$H reaction is plotted as a function of 
$\theta_\pi$ which is the angle between the outgoing $\pi^+$ 
and the three--momentum transfer $\vec{q}$. The energy transfer was chosen
to be $\omega_{\text{lab}} = 300$ MeV and the neutron scattering angle
is $\theta_n = 0^o$. One recognizes that the angular distribution 
is strongly forward peaked. This means that most of the coherent pions
are emitted into the direction of the three--momentum transfer.

\begin{figure}
  \caption{Triple differential cross section for the $^2$H$(p,n\pi^+)^2$H 
     reaction in the spin--longitudinal (LO) and the spin--transversal 
     (TR) channel. $M_i$ ($M_f$) refers to 
     the spin--projection of the deuteron in the initial (final) state,
     hence the dashed curves labeled $M_i=M_f$ show a calculation where 
     spin flips of the deuteron have been excluded.}
  \label{fig14}
  \begin{center}
    \epsfig{file=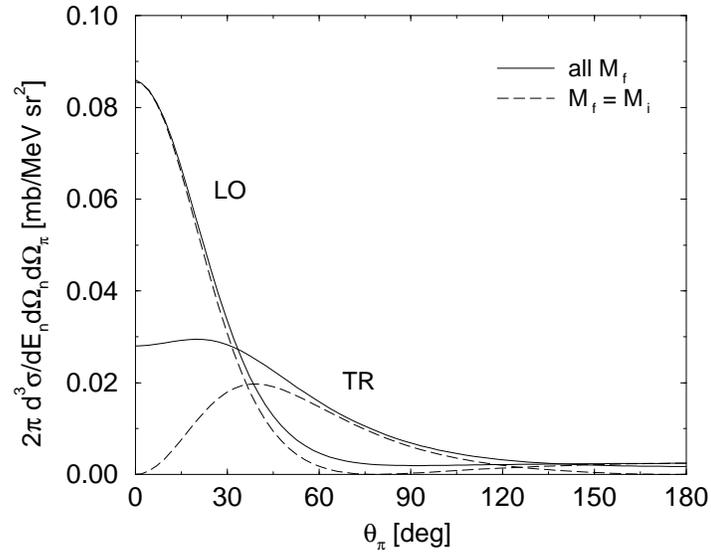, height=8.2cm}
  \end{center}
\end{figure}

The contributions of the spin--longitudinal (LO) and the spin--transverse
(TR) channel to the differential cross section are shown separately in
Fig.\ \ref{fig14}. One can understand some of the main features of
these angular distributions by analyzing the spin structure of the
transition operators. 

For the LO channel, the product of the de--excitation
and excitation operators turns out to be
\begin{equation}
  \label{eq:spin1}
  (\vec{S} \cdot \vec{p_\pi}) (\vec{S}^\dagger \cdot \vec{q} \, ) = 
  \frac{2}{3} \, \vec{p_\pi} \cdot \vec{q} -
  \frac{i}{3} \, \vec{\sigma} \cdot ( \vec{p_\pi} \times \vec{q} ).
\end{equation}
The first term on the r.h.s.\ of Eq.\ (\ref{eq:spin1}) gives rise to a 
factor $q p_\pi \cos \theta_\pi$ in the transition amplitude 
and thus to a factor $\cos^2 \theta_\pi$ in the cross section.
This factor has a maximum at $\theta_\pi = 0^o$ and thus
partially explains the strongly forward peaked LO contribution 
in Fig.\ \ref{fig14}. There are, however, additional angular dependent
factors such as the kinematical phase space factor and the 
overlap integral with the outgoing pion wave in Eq.\ (\ref{eq31}).
Those factors become larger as $\theta_\pi$ gets smaller and therefore 
pronounce the forward peaking of the angular distribution even more. 
All reaction events where the spin 
of the deuteron is flipped are produced by the second term
on the r.h.s.\ of Eq.\ (\ref{eq:spin1}). At small scattering angles
this cross section contribution is proportional to $\sin^2 \theta_\pi$
and thus vanishes at $\theta_\pi=0^o$.

We would like to mention that the angular distribution of the LO component 
is very similar to that of the pion elastic scattering 
\cite{koerfgen94,wilhelm96}. In fact, one may view the coherent pion 
production as a kind of elastic scattering process, in which an initially
off--mass--shell pion with the momentum $\vec{q}$ is converted into
an on--mass--shell pion by the multiple scattering in the deuteron.
This conversion process is possible because the deuteron as a whole
can provide the extra recoil momentum needed to put the pion on its
mass shell.

For the TR channel, the de--excitation and excitation operators yield
a spin structure
\begin{equation}
  \label{eq:spin2}
  (\vec{S} \cdot \vec{p_\pi}) (\vec{S}^\dagger \times \vec{q} \, ) = 
  \frac{2}{3} \, \vec{p_\pi} \times \vec{q} -
  \frac{i}{3} \, ( \vec{\sigma} \times \vec{p_\pi} ) \times \vec{q} .
\end{equation}
As before, the first term on the r.h.s.\ of Eq.\ (\ref{eq:spin2}) 
is connected with non--spin--flip events while the second term induces 
spin--flips of the deuteron. 
The angular distribution of the non--spin--flip part is proportional
to $| \vec{p_\pi} \times \vec{q} \, |^2 = (p_\pi q \sin \theta_\pi)^2 $. 
This factor vanishes for $\theta_\pi = 0^o$ and peaks for 
$\theta_\pi = 90^o$. The additional angular dependent factors as discussed
before lead to a more forward peaked distribution. 
As can be seen from Fig.\ \ref{fig14}, the TR non--spin--flip component 
has its maximum at about $\theta_\pi \approx 40^o$ and is zero at
$\theta_\pi = 0^o$. 
The full TR angular distribution, however, has a different shape
due to the spin--flip contributions to the cross section. 
This demonstrates the importance of the second term on the r.h.s.\ 
of Eq.\ (\ref{eq:spin2}) which does not vanish 
but rather reaches its maximum at $\theta_\pi = 0^o$. 
The overall result is quite a flat angular distribution at small 
pion angles and a fall off at higher angles which is less steep 
as compared to the LO component. This behavior of the TR component
is very similar to the observed angular distribution in pion 
photoproduction ($\gamma$,$\pi$) reactions on the deuteron,
see e.g.\ Ref.\ \cite{ericson88}.


\section{Summary and Conclusions}


In summary we have shown that $(p,n)$ charge exchange reactions on a 
deuteron target 
provide an excellent tool to investigate the $\Delta N$ interaction. 
They probe both the LO and the TR response function in a kinematical region 
which is inaccessible to real pion or photon reactions. 
We have presented a coupled channel approach which allows the 
theoretical treatment of the interacting $\Delta N$ system. 
By assuming that the $\Delta$ resonance is excited dominantly 
in direct interaction with the projectile, we modified the coupled 
equations and showed how to solve for the correlated $\Delta N$ wave 
function in a very efficient way with the Lanczos method. 
For the $\Delta N$ potential, a meson exchange model was adopted which 
includes $\pi$, $\rho$, $\omega$, and $\sigma$ exchange currents.

Results of numerical analysis have been shown for inclusive and exclusive 
cross sections in the $\Delta$ energy region. 
In the coherent pion production $^2$H$(p,n\pi^+ )^2$H, a strong shift of the
$\Delta$ peak position is observed. We have shown that this lowering
in excitation energy is due to the strongly attractive correlations in
the LO spin--isospin channel. This attraction emerges mainly from the 
energy dependent $\pi$ exchange potential. 
Furthermore, we calculated the angular distribution of the coherent pion 
component and found it to be strongly forward 
(in the direction of the momentum transfer) peaked. 

The theoretical calculation for the inclusive $^2$H$(p,n)$ cross section 
is in fairly good agreement with experimental data. 
The model describes both the $\Delta$ peak and the quasi--elastic peak 
very well.
In the dip--region, theory and measurement agree only in the LO channel 
while experimental data are underestimated in the TR channel. 
The enhancement in the TR channel is most probably due to rescattering 
effects (i.e. purely mesonic exchange currents). If such effects could
explain the shortcomings of the present model, they would also demonstrate
the validity limits of the impulse approximation. 
Therefore rescattering effects are of special interest and need further 
investigation. 
In addition, our model may be improved by including higher nucleon 
($N^*$) resonances in the model space. 
Both rescattering effects and $N^*$ resonances will be subject of 
future studies.

\section*{Acknowledgments}

We are very grateful to D.~L. Prout for making the experimental data
available to us. We also thank B.\ K\"orfgen for many helpful discussions.
This work was supported in part by the Studienstiftung des deutschen
Volkes.

\vspace*{0.6cm}


\appendix


\section{The effective projectile -- target--nucleon interaction}
\label{app:1}

In the present paper, we approximate the spin--isospin part 
of $t_{NN,NN}$ in the $NN$ c.m. frame by
\begin{equation}
  \label{eqA1}
  t_{NN,NN}(s,t) = 
  [ \alpha \: (\vec{\sigma}_i\cdot\hat{q}) (\vec{\sigma}_j\cdot\hat{q})
   +\beta  \: (\vec{\sigma}_i\times\hat{q}) (\vec{\sigma}_j\times\hat{q}) ]
  \vec{\tau}_i\cdot\vec{\tau}_j ,
\end{equation}
where the unit vector $\hat{q}$ is connected to the initial
and final nucleon momenta $\vec{\kappa}$ and $\vec{\kappa}\,'$ by
$\vec{q}=\vec{\kappa}-\vec{\kappa}\,'$. The coefficients $\alpha$
and $\beta$ are functions of the Mandelstam variables $s$ and $t$
and describe the strength of LO and TR spin excitations, respectively. 
They are determined from experimental $pn \rightarrow np$ scattering 
data \cite{gaarde91} which can be fitted with
\begin{equation}
  \alpha = c_{\alpha} 
  \left( \frac{\Lambda^2_{\alpha1}+t}{\Lambda^2_{\alpha1}-t} \right) 
  \left( \frac{\Lambda^2_{\alpha2}-m_{\pi}^2}{\Lambda^2_{\alpha2}-t} \right) ,
  \qquad
  \beta = c_{\beta} 
  \left( \frac{\Lambda^2_{\beta}-m_{\pi}^2}{\Lambda^2_{\beta}-t} \right) ,
\end{equation}
where
$ c_{\alpha} = 208 $ MeV fm$^3$, $ c_{\beta} =  178 $ MeV fm$^3$,
$ \Lambda_{\alpha1} = 148 $ MeV, $ \Lambda_{\alpha2} = 460 $ MeV, and
$ \Lambda_{\beta} = 342 $ MeV. 

For the $t_{NN,N\Delta}$ transition operator we assume 
in complete analogy the following form,
\begin{equation}
  \label{eqA2}
  t_{NN,N\Delta}(s,t) =
  [ \gamma \: (\vec{\sigma}_i\cdot\hat{q}) (\vec{S}_j^\dagger\cdot\hat{q})
   +\gamma\,' \: (\vec{\sigma}_i\times\hat{q}) 
                 (\vec{S}_j^\dagger\times\hat{q}) ]
   \vec{\tau}_i\cdot\vec{T}_j^\dagger .
\end{equation}
Again, the coefficients $\gamma$ and $\gamma\,'$ are adjusted to fit 
experimental data, here of the $pN \rightarrow n\Delta$ reactions
\cite{prout}, yielding 
\begin{equation}
  \gamma = \gamma\,' = c_{\gamma} 
  \left( \frac{\Lambda_{\gamma}^2-m_{\pi}^2}{\Lambda_{\gamma}^2-t} \right) , 
\end{equation}
where $c_{\gamma} = 488$ MeV fm$^3$ and $\Lambda_{\gamma} = 650$ MeV.
The fact that data are explained with $\gamma = \gamma\,'$ means that the 
ratio LO/TR is 1/2.

The assumed operators are essentially of $\delta$--function type;
the only momentum dependence comes from the vertex form factors.
In spite of their simplicity, $t_{NN,NN}$ and $t_{NN,N\Delta}$ can reproduce 
not only the cross sections but also the spin observables from the reactions 
with the proton target \cite{udagawa94}.


\section{Evaluation of matrix elements}

All matrix elements which appear in this work are evaluated by using partial
wave expansions and standard tensor operator techniques, see e.g.\ Ref.\ 
\cite{edmonds57}. In this appendix, we discuss in some detail the
calculation of the radial source function and of the matrix elements for the
$\Delta N$ interaction. The matrix elements of Sec.\ \ref{sec:decomp} are
not given explicitly but can be calculated rather easily within the presented
scheme.

\subsection{Explicit formula for the source functions} 
\label{app:2}

For the $\Delta N$ system, the radial source function is obtained from
Eq.\ (\ref{eq18}) by inversion
\begin{equation}
  \label{eqB1}
  \rho_{SLJM_J} (r) = 
  r \: \langle (SL)JM_J \mid \rho_{\Delta} \rangle = 
  r \: \langle (SL)JM_J \mid \hat{\rho} \mid \Psi_d \rangle , 
\end{equation}
where the deuteron wave function is given as
\begin{equation}
  \label{eqB2}
  \mid \Psi_d \rangle = \sum_{l_d=0,2} \frac{1}{r} \phi_{l_d} (r) 
  \mid (1 l_d) 1 M_d \rangle .
\end{equation}
The hadronic transition operator $\hat{\rho}$ was defined in 
Eq.\ (\ref{eq8}). If we apply the specific form of $t_{NN,N\Delta}$ 
as given in Eq.\ (\ref{eqA2}) and use the tensor operator notation, 
$\hat{\rho}$ may be rewritten as
\begin{equation}
  \label{eqB3}
  \hat{\rho} = ( \text{IF} ) \sum_{lm\nu} \tilde{\rho}_{lm\nu} (r)
  \: (S_1^{\dagger})^{(1)}_{\nu} \, Y_{lm} (\hat r) 
  + (1 \leftrightarrow 2) ,
\end{equation}
where
\begin{equation}
  \tilde{\rho}_{lm\nu} (r) = 4 \pi \, \gamma(s,t) \, 
  (-)^{m_p + \frac{1}{2}} 
  \left( \begin{array}{ccc} 1 & \frac{1}{2} & \frac{1}{2} \\ 
  \nu & m_n & m_p \end{array} \right)
  \langle \frac{1}{2} \mid \mid \vec{\sigma} \mid \mid \frac{1}{2} \rangle
  \, i^l \, j_l(\frac{1}{2} qr) \, Y^*_{lm} (\hat q)
\end{equation}
and (IF) denotes the isospin factor 
\begin{equation}
  (\text{IF}) = \langle \Delta N \mid \langle n \mid 
  \vec{\tau_0} \cdot \vec{T}^{\dagger} \mid p \rangle \mid d \rangle = 
  \delta_{1T} \delta_{1M_T} \frac{2}{\sqrt{3}} .
\end{equation}
Insertion of Eqs.\ (\ref{eqB2},\ref{eqB3}) into Eq.\ (\ref{eqB1}) yields
\begin{eqnarray}
  \label{eqB4}
  \rho_{SLJM_J} (r) &=& ( \text{IF} ) \sum_{lm\nu} \tilde{\rho}_{lm\nu} (r)
  \sum_{l_d} \phi_{l_d} (r) 
  \nonumber \\ && \times
  \langle (SL)JM_J \mid (S_1^{\dagger})^{(1)}_{\nu} Y_{lm} (\hat r)
  \mid (1l_d) 1 M_d \rangle 
  + (1 \leftrightarrow 2) .   
\end{eqnarray}
The calculation of the angular momentum matrix element is straight forward, 
but requires some lengthy spin algebra. The final result is
\begin{eqnarray}
  \label{eqB5}
  && \langle (S L ) J M_J \mid (S_1^{\dagger})^{(1)}_{\nu} Y_{lm} (\hat r)
  \mid (1l_d) 1 M_d \rangle = 
  \nonumber \\ && 
  \sqrt{\frac{3}{4\pi}} \, \hat{J} \hat{L} \hat {l} \hat{l_d} \,
  (-)^{L+1} ( \delta_{S1} + \sqrt{5} \delta_{S2} )
  \sum_{J_i M_i} (-)^{(M_J-M_i)} (2J_i + 1)  
  \nonumber \\ && \times
  \left( \begin{array}{ccc} l & 1 & J_i \\ m & M_d & -M_i \end{array} 
  \right)
  \left( \begin{array}{ccc} J_i & 1 & J \\ M_i & \nu & -M_J \end{array} 
  \right) 
  \left( \begin{array}{ccc} l & l_d & L \\ 0 & 0 & 0 \end{array} 
  \right)
  \left\{ \begin{array}{ccc} L & J_i & 1 \\ 1 & l & l_d \end{array} 
  \right\} 
  \left\{ \begin{array}{ccc} S & 1 & 1 \\ J_i & J & L \end{array} 
  \right\} .
\end{eqnarray}

For the radial source function of the $NN$ system, an equation analogous 
to (\ref{eqB4}) holds with $\vec{S}^{\dagger}$ replaced by $\vec{\sigma}$,
an isospin factor $(\text{IF}) = \sqrt{2}$, and a transition strength 
$\alpha(s,t)$ resp.\ $\beta(s,t)$ instead of $\gamma(s,t)$.


\subsection{Matrix elements of the $\Delta N$ potential}
\label{app:3}

In configuration space, the $\Delta N$ potentials
$V_{\Delta N \rightarrow \Delta N}$ (direct) and 
$V_{\Delta N \rightarrow N \Delta}$ (exchange) 
as well as the transition potential
$V_{\Delta N \rightarrow NN}$ exhibit the following spin--isospin structure,
\begin{equation}
  \label{eqC1}
  V_{ab \rightarrow a'b'} (k_0, \vec{r} \, ) = 
  {\cal F}^{\text{N}} (r) +
  \left[ \vec{\sigma}^{\, aa'}_1 \cdot \vec{\sigma}^{\, bb'}_2 
    {\cal F}^{\text{C}} (r) 
    + S_{12}^{\, aa',bb'} (\hat{r}) {\cal F}^{\text{T}} (r) 
  \right] \vec{\tau}_1^{\, aa'} \cdot \vec{\tau}_2^{bb'} .
\end{equation}
Here, $\vec{\sigma}^{\, aa'}$ and $\vec{\tau}^{\, aa'}$ are the 
appropriate spin and isospin operators with $a,a' \in \{ N,\Delta \}$ and
\begin{equation}
  \label{eqC2} 
  S_{12}^{\, aa',bb'} (\hat{r}) = 
  3 \left( \vec{\sigma}^{\, aa'}_1 \cdot \hat{r} \right) 
    \left( \vec{\sigma}^{\, bb'}_2 \cdot \hat{r} \right)  
  - \vec{\sigma}^{\, aa'}_1 \cdot \vec{\sigma}^{\, bb'}_2 
\end{equation}
is the usual spin tensor operator. In terms of the spin operators used
in Sec.\ \ref{sec:vdn}, we have  
$ \vec{\sigma}^{NN} = \vec{\sigma} $,
$ \vec{\sigma}^{N \Delta} = \vec{S}^{\dagger} $ and  
$ \vec{\sigma}^{\Delta \Delta} = \vec{\Sigma} $,
thus the reduced matrix elements are given by 
\begin{equation}
  \label{eqC3}
  \langle \frac{1}{2} \mid\mid \vec{\sigma}^{NN} 
    \mid\mid \frac{1}{2} \rangle = \sqrt{6} , \qquad
  \langle \frac{3}{2} \mid\mid \vec{\sigma}^{N \Delta} 
    \mid\mid \frac{1}{2} \rangle = 2 , \qquad
  \langle \frac{3}{2} \mid\mid \vec{\sigma}^{\Delta \Delta} 
    \mid\mid \frac{3}{2} \rangle = 2 \sqrt{15} . 
\end{equation}
Equivalent equations hold for the isospin operators $\vec{\tau}^{\, aa'}$.

The functions $\cal F$ of Eq.\ (\ref{eqC1}) are the 
Fourier transformations of the non--spin, spin--central and spin--tensor
part of the interaction, i.e.
\begin{mathletters}
  \label{eqC4}
  \begin{eqnarray}
    {\cal F}^{\text{N}} (r) &=& 
      \frac{1}{2 \pi^2} \int dk \: k^2  j_0 (kr) 
      V^{\text{N}}_{ab,a'b'} (k_0,k) , \\
    {\cal F}^{\text{C}} (r) &=& 
      \frac{1}{2 \pi^2} \int dk \: k^2  j_0 (kr)
      \left[ V^{\text{LO}}_{ab,a'b'} (k_0,k) + 2 V^{\text{TR}}_{ab,a'b'} 
      (k_0,k) \right] , \\
    {\cal F}^{\text{T}} (r) &=& 
      \frac{-1}{2 \pi^2} \int dk \: k^2  j_2 (kr)
      \left[ V^{\text{LO}}_{ab,a'b'} (k_0,k) - V^{\text{TR}}_{ab,a'b'} 
      (k_0,k) \right] .
  \end{eqnarray}
\end{mathletters}
For the potentials given in Sec.\ \ref{sec:vdn}, these expressions may 
easily be found analytically. 
Furthermore, we need the following spin matrix elements
which are calculated with standard techniques, 
\begin{eqnarray}
  \label{eqC6}
  \langle [ ( s_{a'} s_{b'} ) S'L' ] JM_J & & \mid 
  \vec{\sigma}_1^{\, aa'} \cdot \vec{\sigma}_2^{\, bb'}
  \mid [ ( s_{a} s_{b} ) SL ] JM_J \rangle = 
  \nonumber \\ 
  & & (-)^{ s_{a} + s_{b'} + S } \delta_{SS'} \delta_{LL'}
      \left\{ \begin{array}{ccc} s_{a'} & s_{b'} & S \\
                                 s_{b}  & s_{a}  & S \end{array} \right\}
  \langle s_{a'} \mid\mid \sigma^{aa'} \mid\mid s_{a} \rangle   
  \langle s_{b'} \mid\mid \sigma^{bb'} \mid\mid s_{b} \rangle , 
  \\
  \label{eqC7}
   \langle [ ( s_{a'} s_{b'} ) S'L' ] JM_J & & \mid 
  S_{12}^{\, aa',bb'} (\hat{r}) \mid [ ( s_{a} s_{b} ) SL ] JM_J \rangle = 
  \nonumber \\ && 
  \sqrt{30} (-)^{(S+J)} \hat{S} \hat{S'} \hat{L} \hat{L'} 
  \left( \begin{array}{ccc} L & L' & 2 \\ 0 & 0 & 0 \end{array} \right)
  \left\{ \begin{array}{ccc} S' & S & 2 \\ L & L' & J \end{array} \right\}
  \left\{ \begin{array}{ccc} s_{a'} & s_{b'} & S' \\
                             s_{a}  & s_{b}  & S \\
                             1      & 1      & 2 \end{array} \right\} 
  \nonumber \\ && \times
  \langle s_{a'} \mid\mid \sigma^{aa'} \mid\mid s_{a} \rangle   
  \langle s_{b'} \mid\mid \sigma^{bb'} \mid\mid s_{b} \rangle . 
\end{eqnarray}
From Eqs.\ (\ref{eqC3}) -- (\ref{eqC7}) one can build up the complete 
expressions for the matrix elements $V_{nn'} (r)$ 
as defined in Eq.\ (\ref{eq19}), i.e.
\begin{equation}
  V_{nn'} (r) =  \langle [ ( s_{a'} s_{b'} ) S'L' ] JM_J \mid 
  \langle (\tau_{a'} \tau_{b'}) 11 \mid V_{ab \rightarrow a'b'} 
  (k_0, \vec{r})
  \mid (\tau_{a} \tau_{b}) 11 \rangle 
  \mid [ ( s_{a} s_{b} ) SL ] JM_J \rangle .
\end{equation}







\vspace*{1cm}

\begin{table}
  \caption{Parameters used in the meson exchange model for the $\Delta N$ 
    interaction.}
  \begin{tabular}{cccccc}
    & $f^2_{\alpha NN} /4\pi$ & $f^2_{\alpha N \Delta} /4\pi$ 
    & $f^2_{\alpha \Delta \Delta} /4\pi$ & $\Lambda_\alpha$ [GeV] 
    & $m_\alpha$ [MeV] \\
    \tableline
    $\pi$ & 0.08 & 0.32 & 0.0032 & 1.1 & 138 \\
    $\rho$ & 5.4 & 21.6 & 0.216 & 1.4 & 770 \\
    $\omega$ & 8.1 \tablenotemark[1] & --- & 8.1 \tablenotemark[1] 
    & 1.7 & 783 \\
    $\sigma$ & 5.7 \tablenotemark[1] & --- & 5.7 \tablenotemark[1] 
    & 1.4 & 570 \\
  \end{tabular}
  \tablenotetext[1]{$g^2_{\alpha NN}/4\pi$ resp.\ 
    $g^2_{\alpha \Delta \Delta}/4\pi$ is given.}
\label{tab:parameter}
\end{table}


\newpage

\newpage

\newpage

\newpage

\newpage

\newpage

\newpage

\newpage

\end{document}